\documentclass[12pt]{iopart}

\usepackage[dvips]{graphicx}

\usepackage[round,authoryear]{natbib}
\bibpunct{(}{)}{,}{a}{}{,}

\begin{document}

\title[Cell inactivation by protons and light ions]{Detailed analysis of the cell-inactivation mechanism by accelerated protons and light ions}

\author{Pavel Kundr\'{a}t}

\address{Institute~of~Physics, Academy~of~Sciences~of~the~Czech~Republic,\\
Na~Slovance~2, CZ-182~21~Praha~8, Czech~Republic\\
E-mail: Pavel.Kundrat@fzu.cz}

\begin{abstract}
Published survival data for V79 cells irradiated by monoenergetic protons, helium-3, carbon, and oxygen ions and for CHO cells irradiated by carbon ions have been analyzed using the probabilistic two-stage model of cell inactivation. Three different classes of DNA damages formed by traversing particles have been distinguished, namely severe single-track damages which might lead to cell inactivation directly, less severe damages where cell inactivation is caused by their combinations, and damages of negligible severity that can be repaired easily. Probabilities of single ions to form these damages have been assessed in dependence on their linear energy transfer (LET) values.

Damage induction probabilities increase with atomic number and LET. While combined damages play crucial role at lower LET values, single-track damages dominate in high-LET regions. The yields of single-track lethal damages for protons have been compared with the Monte Carlo estimates of complex DNA lesions, indicating that lethal events correlate well with complex DNA double-strand breaks. The decrease in the single-track damage probability for protons of LET above approx.\ 30 keV/$\mu$m, suggested by limited experimental evidence, is discussed, together with the consequent differences in the mechanisms of biological effects between protons and heavier ions. Applications of the results in hadrontherapy treatment planning are outlined.
\end{abstract}

\maketitle

%%%%%%%%%%%%%%%%%%%%%%%%%%%%%%%%%%%%%%%%%%%%%%%%%%%%%%%%%%%%%%%%%%%%%%%%%%%%%%%%%%%%
\section{Introduction}
\label{sec:Introduction}
Hadron radiotherapy, based on irradiating tumours by beams of accelerated protons and other ions, is expected to significantly increase the cure rate of cancer in near future; compare e.g.\ \citep{MedAustron, CNAO, Brahme - Design, Heidelberg}. The well-known rationale for using proton beams in radiotherapy lies in the characteristic pattern of their energy deposition to matter, the Bragg peak, leading to the possibility of achieving highly conformal dose distributions \citep{Wilson, Lomax1, Lomax2}. Beams of light ions possess additional advantages: They are characterized by higher linear energy transfer (LET) values, leading to enhanced relative biological effectiveness (RBE) and diminishing oxygen enhancement ratio (OER), e.g.\ \citep{ion-rationale}. They are therefore expected to be especially suitable for curing tumours resistant to conventional radiotherapy; compare e.g.\ \citep{ion-rationale, chordomas}. Another advantage consists in the possibility of online dose monitoring by positron emission tomography, PET \citep{PET}. Furthermore, recent clinical studies indicate that unconventional fractionation schemes might be used for selected tumours, reducing the overall treatment time to a few sessions only \citep{Tsujii - PTCOG Paris 2004}.

Several dedicated hadrontherapy facilities have been launched recently, compare \citep{Particles}. In most of them only proton beams are available; however, in several centres, existing or being built, protons as well as light ions up to carbon or oxygen may be used. For the choice of an optimal ion in given clinical situations, analyses of energy loss, scattering and fragmentation phenomena for different ions and comparisons of their dose distributions are necessary \citep{Brahme - Design, TRiP1, fragmentation}. These physical analyses have to be complemented then by biology-oriented studies, analyzing the mechanisms of biological effects for different ions using available data and adequate models. 

However, existing hadrontherapy treatment planning approaches have been based on detailed description of the physical processes only, and have not addressed the biological phase of the underlying radiobiological mechanism in detail. E.g.\ the treatment planning procedure used at HIMAC (Chiba, Japan) is based on interpolating the $\alpha, \beta$ coefficients of the linear-quadratic (LQ) fits to experimental survival data, and on the similarity in biological effects between carbon and neutron beams of similar LET values \citep{Kanai 1997, Kanai 1999}; compare also \citep{Kagawa 2002}. In the approach used in clinical hadrontherapy applications at GSI (Darmstadt, Germany), the local effect model (LEM), the effect of ion tracks on a microscopic scale is assumed to be equal to that of correspondingly high photon doses (compare \citep{TRiP2}); again, the underlying biological processes of damage induction and repair have not been taken into account explicitly. On the contrary, the present paper summarizes the results of a detailed biology-oriented analysis of cell inactivation data for protons, helium-3, carbon and oxygen ions. Experimental data gathered by several groups \citep{Belli, Perris, Goodhead, Folkard96, Wilma, Kiefer} have been analysed with the help of a detailed probabilistic radiobiological model \citep{PhysMedBiol2005}.

%%%%%%%%%%%%%%%%%%%%%%%%%%%%%%%%%%%%%%%%%%%%%%%%%%%%%%%%%%%%%%%%%%%%%%%%%%%%%%%%%%%%%

%%%%%%%%%%%%%%%%%%%%%%%%%%%%%%%%%%%%%%%%%%%%%%%%%%%%%%%%%%%%%%%%%%%%%%%%%%%%%%%%%%%%%
\section{Probabilistic two-stage model of radiobiological effects}
\label{sec:ProbabilisticRadiobiologicalModel}

%%%%%%%%%%%%%%%%%%%%%%%%%%%%%%%%%%%%%%%%%%%%%%%%%%%%%%%%%%%%%%%%%%%%%%%%%%%%%%%%%%%%%
The processes running in a cell after irradiation (under conditions usual in radiotherapy) proceed in two distinct, subsequent phases. The first stage includes processes running immediately after the impact of individual ionizing particles, i.e.\ energy transfer events, radical formation and diffusion, chemical reactions and formation of DNA damages. The other stage is formed then by the subsequent cellular response which follows as a reaction to the effect of all ionizing particles having hit a cell nucleus or chromosomal system at a given dose; this stage includes the processes of damage repair or misrepair and further biological processes leading to cell survival or inactivation. Distinguishing these two phases represents the basis of the probabilistic two-stage model of biological effects of ionizing particles.

In the full scheme of the model, the following characteristics are taken into account: (i) the stochastic distribution of particle tracks over the irradiated cell population, (ii) the distribution of transferred energy, (iii) the probability of single traversing particles to induce DNA damages of different severity, (iv) saturation or synergetic combinations of individual damages, and (v) the effects of cellular repair systems. The model scheme, presented in \citep{PhysMedBiol2005}, allows to describe different classes of survival curves, including also low-dose hypersensitivity phenomena, on the basis of detailed characteristics of the mentioned processes, especially the effects of damage induction and repair processes. However, in the following comparison of the effects of different light ions that are expected to be used in hadrontherapy, a simplified model scheme will be used, taking into account only DNA damages that are practically unrepairable.

\begin{figure}[!htb]
	\centering
		\includegraphics[angle=0, width=0.60\textwidth]{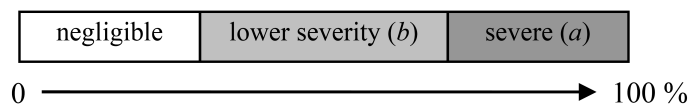}
	\caption{Schematic illustration of the damage classification according to damage complexity.}
	\label{fig:classification}
\end{figure}

The cell inactivation probability at applied dose $D$ is given by
\begin{equation} \label{sumaPkqk}
	s(D) = \sum_k P_k(D) \: q_k \ .
\end{equation}
Here, the number $k$ ($k\geq 0$) of primary particles (particle tracks) traversing individual cell nuclei is given by Poisson distribution
\begin{equation} \label{Poisson}
	P_k (D) = \frac{(hD)^k}{k!} \exp(-hD) \ ,
\end{equation}
the average number per 1 Gy, $h$, being proportional to the geometric cross-section of the nuclei, $\sigma$. The survival probability after $k$ particles have traversed the nucleus has been denoted by $q_k$. It can be calculated by
\begin{equation} \label{qk}
	q_k = 1 - 	\sum_{i=0}^k (^k_{\, i}) a^i (1-a)^{k-i}      
				\sum_{j=0}^{k-i} (^{k-i}_{\, \ j}) b^j (1-b)^{k-i-j}
				[1-r_{ij}^{ab}] \ , 
\end{equation}
where combinatorial numbers $(^k_{\, i}) = \frac{k!}{i!(k-i)!}$. Here $a$ and $b$ stand for the induction probability of different damages and $r_{ij}^{ab}$ for repair probabilities of their different combinations; $r_{00}^{ab} \equiv 1 \equiv r_{01}^{ab}$ (compare below). Three different classes of damage formed by individual particles have been distinguished:
\begin{enumerate}
	\item Severe damage formed by a single track that is capable of causing cell inactivation even if this is the only damage to the cell ("single-track" or "single-particle induced" damage). The probability of inducing such a damage, per track, has been denoted by $a$.
	\item Damages of lower severity, at least two of which must combine to be lethal ("combined" or "two-track" damages). The probability of their formation (per track) has been denoted by $b$.
	\item Lesions that do not represent any significant challenge to the cell and its repair systems. Such negligible lesions have not been included in Eq.\ (\ref{qk}).
\end{enumerate}
The above classification of DNA damage is illustrated in Figure~\ref{fig:classification}.

Eq.\ (\ref{qk}) represents the general formula for cell inactivation after a passage of $k$ particles. The effects of cell repair systems have been included in repair probabilities $r^{ab}_{ij}$. In the present analysis, however, only unrepairable damages have been considered, for which $r^{ab}_{ij} \equiv 0$. In other words, the probabilities $a$ and $b$ derived here represent the lowest estimates of damage induction at corresponding energy transfers (LET values), since unrepairable damages form, essentially, subsets of all (both repairable and unrepairable) damages in the above classification (Eq.\ (\ref{qk}), Figure~\ref{fig:classification}).

Considering unrepairable damages only, cell inactivation follows, then, if at least one single-track or at least two combined-type damages have been formed by the $k$ particles traversing the nucleus. Indeed, for $r^{ab}_{ij} \equiv 0$, Eq.\ (\ref{qk}) may be simplified to
\begin{equation} \label{qk2}
	q_k = (1-a)^k - 	(1-a)^k  \sum_{j=2}^k (^k_{\, j})  b^j (1-b)^{k-j}  
	= (1-a)^k [(1-b)^k + kb(1-b)^{k-1}]
\end{equation}
(with $q_0 = 1$ and $q_1 = 1-a$), 
i.e., the cell survives if no single-track and not more than a single $b$-type damage have been formed by the $k$ particles.
\footnote{The corresponding Eqs.\ (9-15) in \citep{PhysMedBiol2005} represented simplified versions of formulas derived from Eq.\ (\ref{qk}), as combined damages $b$ were taken into account in a simplified manner only: The synergetic effects were represented by terms $(1-b^2)^{k(k-1)/2}$ only, yielding e.g.\ $q_k = (1-a)^k (1-b^2)^{k(k-1)/2}$ after neglecting the repair probabilities. By comparing this formula with Eq.\ (\ref{qk2}) it is clear that the role of combined damages was slightly underestimated, especially in the region of higher particle numbers. I.e.\ the analyses based on precise formulas derived from Eq.\ (\ref{qk}) yield slightly higher combined-damage induction probabilities as compared to the simplified formulas used in \citep{PhysMedBiol2005}. This fact, however, does not affect the conclusions drawn in that paper, especially those concerning the role of single-particle and combined damage induction and repair processes with respect to the shapes of survival curves.}

%%%%%%%%%%%%%%%%%%%%%%%%%%%%%%%%%%%%%%%%%%%%%%%%%%%%%%%%%%%%%%%%%%%%%%%%
\section{Analysis of experimental data}
\label{sec:AnalysisOfExperimentalData}

Survival data for V79 cells irradiated by protons at 0.57~--~7.4~MeV \citep{Belli, Folkard96, Perris, Goodhead}, helium-3 at 3.4~--~6.9~MeV \citep{Folkard96}, carbon at 2.4~--~266~MeV/u \citep{Wilma} and oxygen ions at 1.9~--~396~MeV/u \citep{Kiefer} have been analysed using the model reviewed in the preceding section. The list of particle energies, LET values and ranges in water is given in Table~\ref{tab:energies}.

%%%%%%%%%%%%%%%%%%%%%%%%%%%%%%%%%%%%%%%%%%%%%%%%%%%%%%%%%%%%%%%%%%%
\begin{table}
\caption{\label{tab:energies}Physical parameters of particles in the analyzed data sets. Energy, LET values and ranges as given by the authors, or calculated using PSTAR tables \citep{PSTAR} or SRIM-2003 \citep{SRIM}.}
\begin{indented}
\item[]\begin{tabular}{@{}lllll}
\br
&Energy&LET&Range in water&Reference\\
&[MeV/u]&[$\mathrm{keV/\mu m}$]&[mm]&\\
\mr
p&7.4&5.8&0.72&\citet{Perris}\\
&5.01&7.7&0.37&\citet{Belli}\\
&3.66&10.1&0.21&\citet{Folkard96}\\
&3.20&11.0&0.17&\citet{Belli}\\
&3.0&11.7&0.15&\citet{Perris}\\
&1.83&17.8&0.065&\citet{Folkard96}\\
&1.41&20.0&0.043&\citet{Belli}\\
&1.4&20.3&0.042&\citet{Goodhead}\\
&1.16&23.0&0.040&\citet{Goodhead}\\
&1.07&27.6&0.027&\citet{Folkard96}\\
&0.76&30.5&0.016&\citet{Belli}\\
&0.64&34.6&0.012&\citet{Belli}\\
&0.57&37.8&0.010&\citet{Belli}\\
\mr
$^3\mathrm{He}$&2.30&58.9&0.074&\citet{Folkard96}\\
&1.39&88.3&0.033&\\
&1.13&105.8&0.024&\\
\mr
C&266.4&13.7&140&\citet{Wilma}\\
&190.7&16.8&80&\\
&76.9&32.4&15.9&\\
&18.0&103&1.2&\\
&11.0&153.5&0.5&\\
&5.4&275.1&0.15&\\
&4.2&339.1&0.1&\\
&2.4&482.7&0.045&\\
\mr
O&396&18&202&\citet{Kiefer}\\
&88&46&15.2&\\
&10.7&238&0.36&\\
&1.9&754&0.030&\\
\br
\end{tabular}
\end{indented}
\end{table}
%%%%%%%%%%%%%%%%%%%%%%%%%%%%%%%%%%%%%%%%%%%%%%%%%%%%%%%%%%%%%%%%%%%

To facilitate a systematic analysis of the given data sets, the damage induction probabilities for a given ion have been considered in dependence on LET value $\lambda$ instead of deriving their values for individual measured survival curves independently. These LET dependences have been represented by flexible test functions involving a low number of auxiliary parameters, $a_{0-2}$ and $b_{0-2}$:
\begin{equation} \label{ab}
	a(\lambda) = a_0 (1 - \exp(-(a_1 \lambda)^{a_2})) \ , \qquad
	b(\lambda) = b_0 (1 - \exp(-(b_1 \lambda)^{b_2})) \ .
\end{equation}
Parameter values have been determined using standard optimization methods; mainly the SIMPLEX and MIGRAD methods implemented in the MINUIT multivariate minimization tool~\citep{MINUIT} have been applied to dedicated computer codes written in FORTRAN. Weighted least-square method has been used to construct the objective function, $\chi^2$, as described previously \citep{PhysMedBiol2005}.

\begin{figure}[!htb]
	\centering
		\includegraphics[trim=0cm 0.4cm 0cm 1.2cm, clip, angle=-90, width=0.32\textwidth]{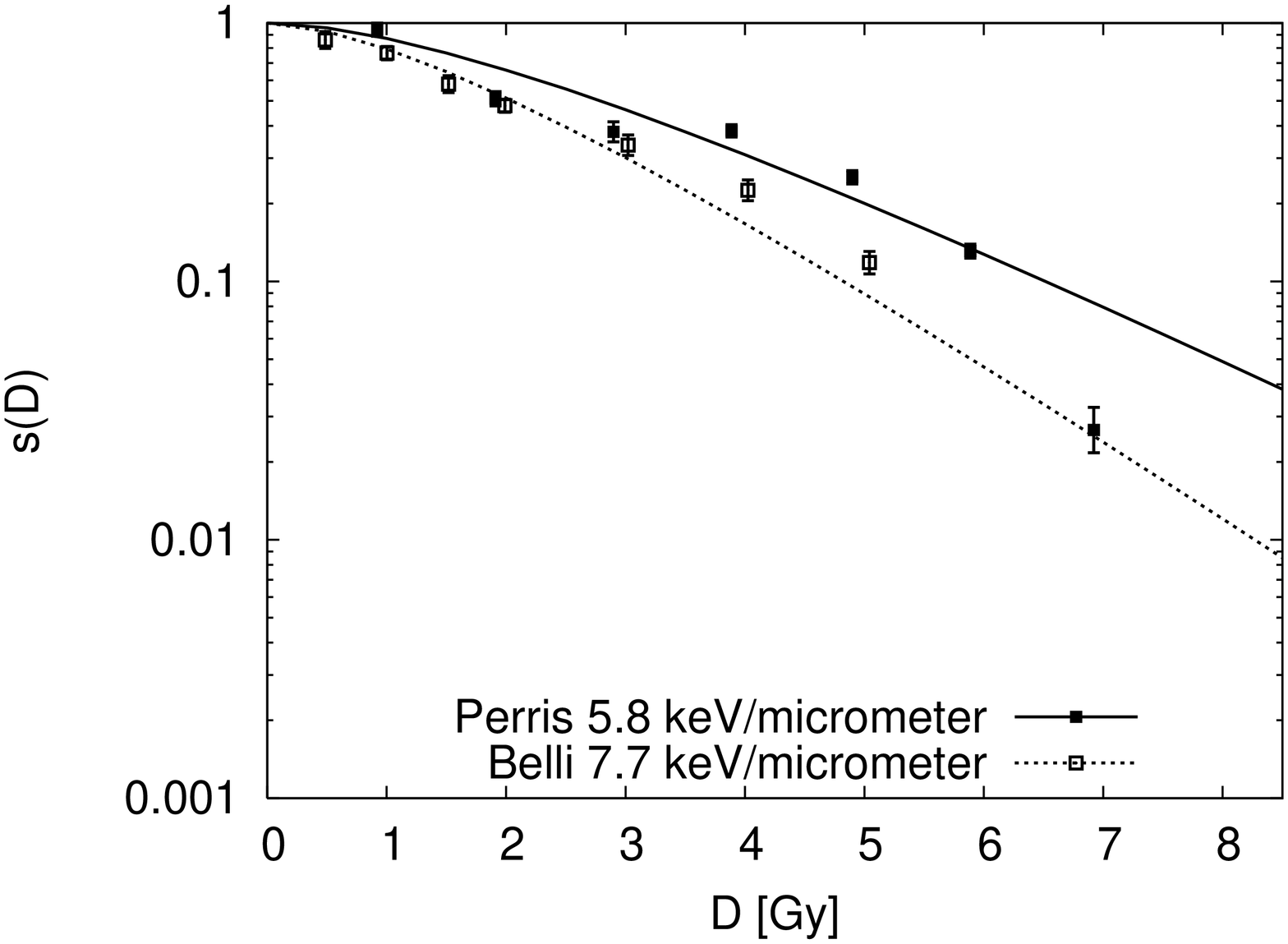}
		\includegraphics[trim=0cm 0.4cm 0cm 1.2cm, clip, angle=-90, width=0.32\textwidth]{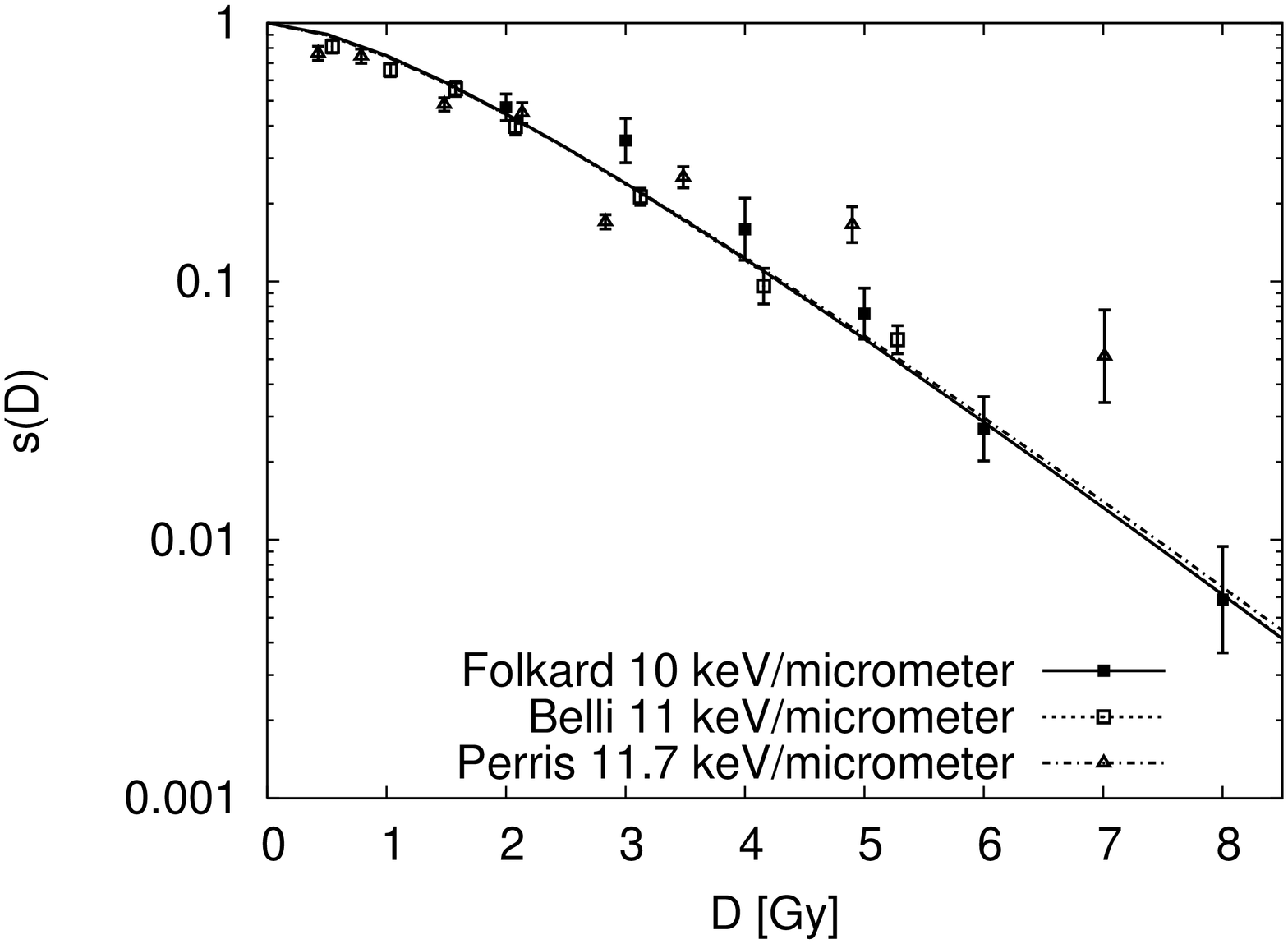}
		\includegraphics[trim=0cm 0.4cm 0cm 1.2cm, clip, angle=-90, width=0.32\textwidth]{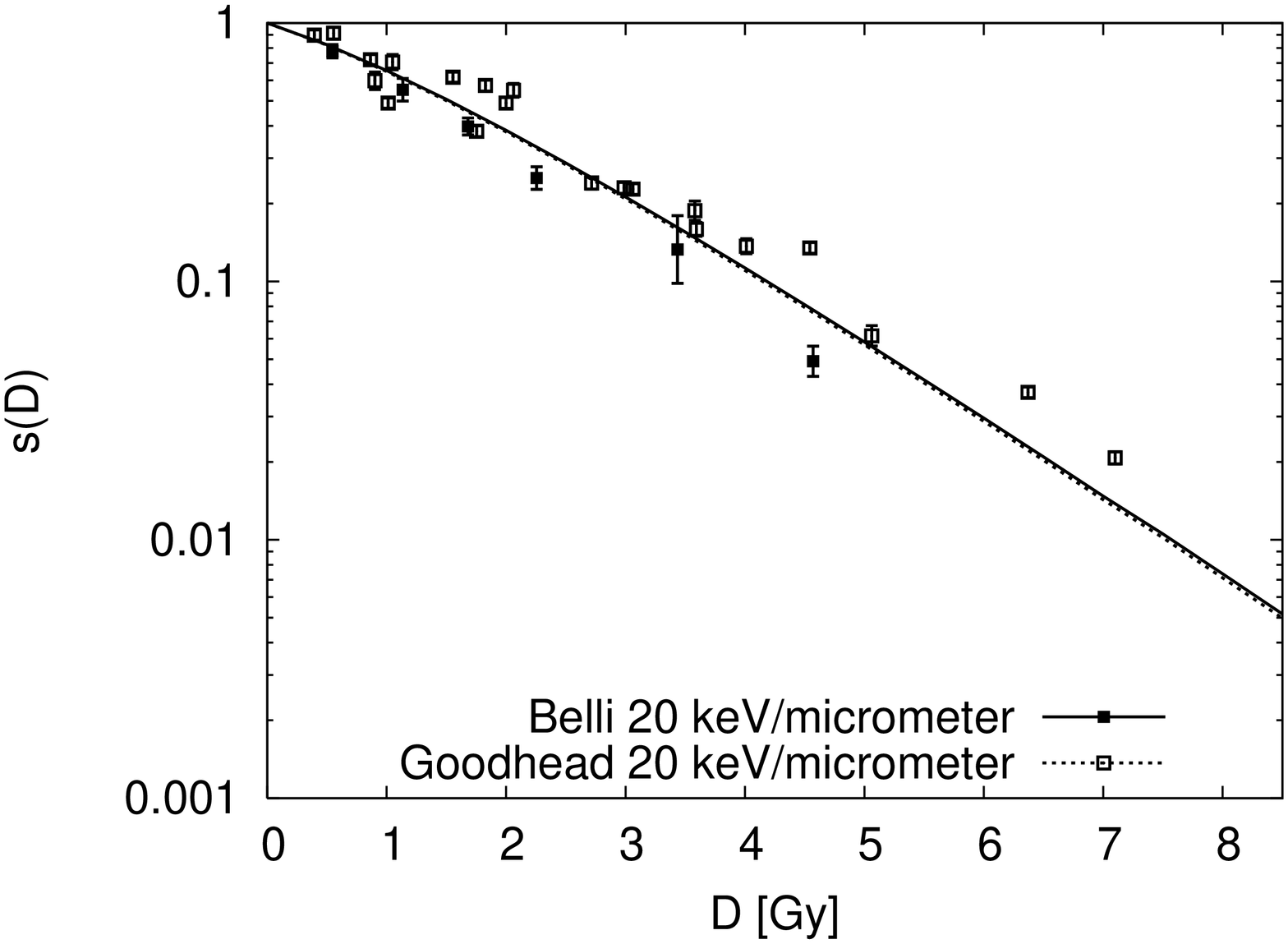}
		\includegraphics[trim=0cm 0.4cm 0cm 1.2cm, clip, angle=-90, width=0.32\textwidth]{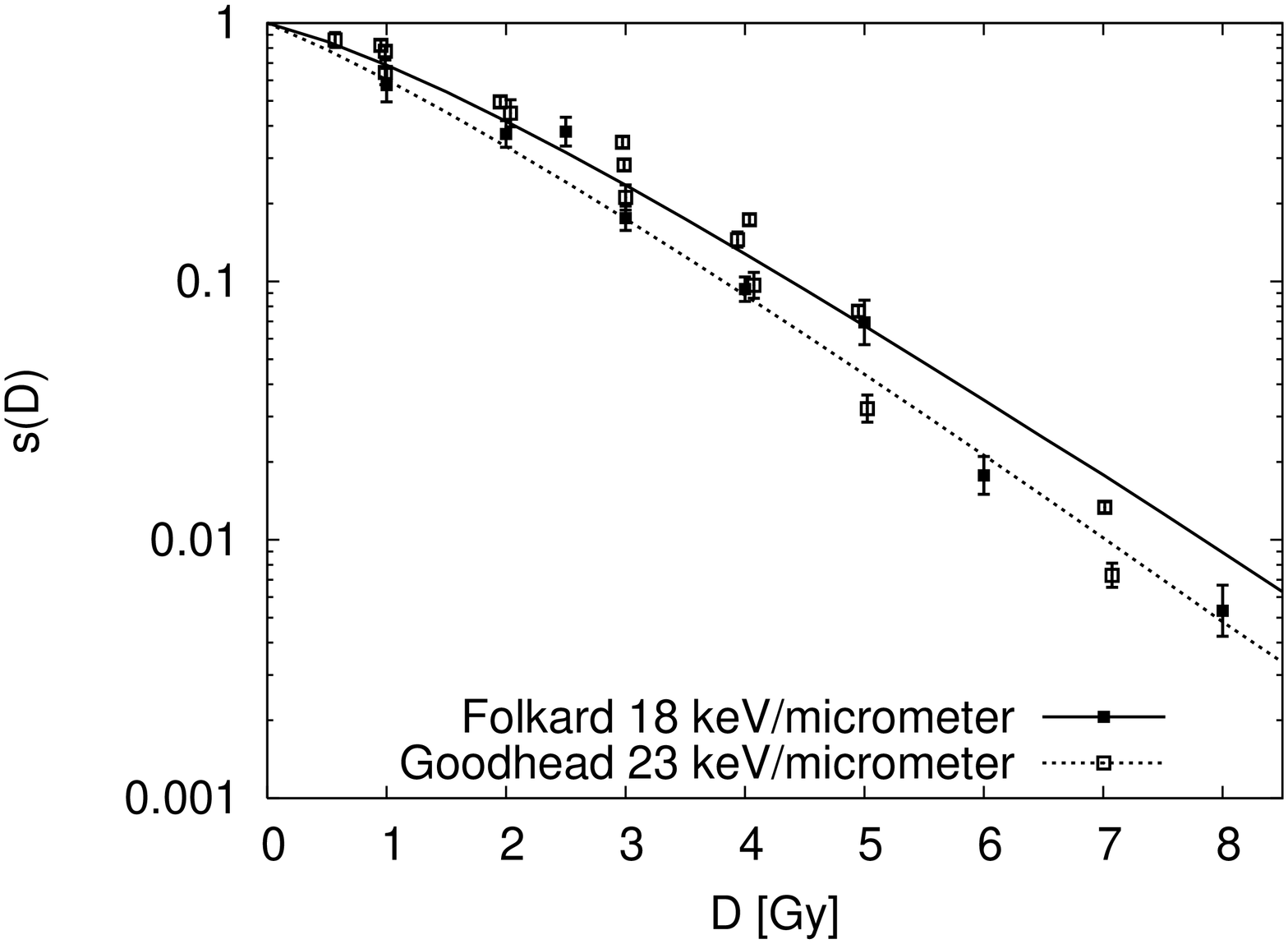}
		\includegraphics[trim=0cm 0.4cm 0cm 1.2cm, clip, angle=-90, width=0.32\textwidth]{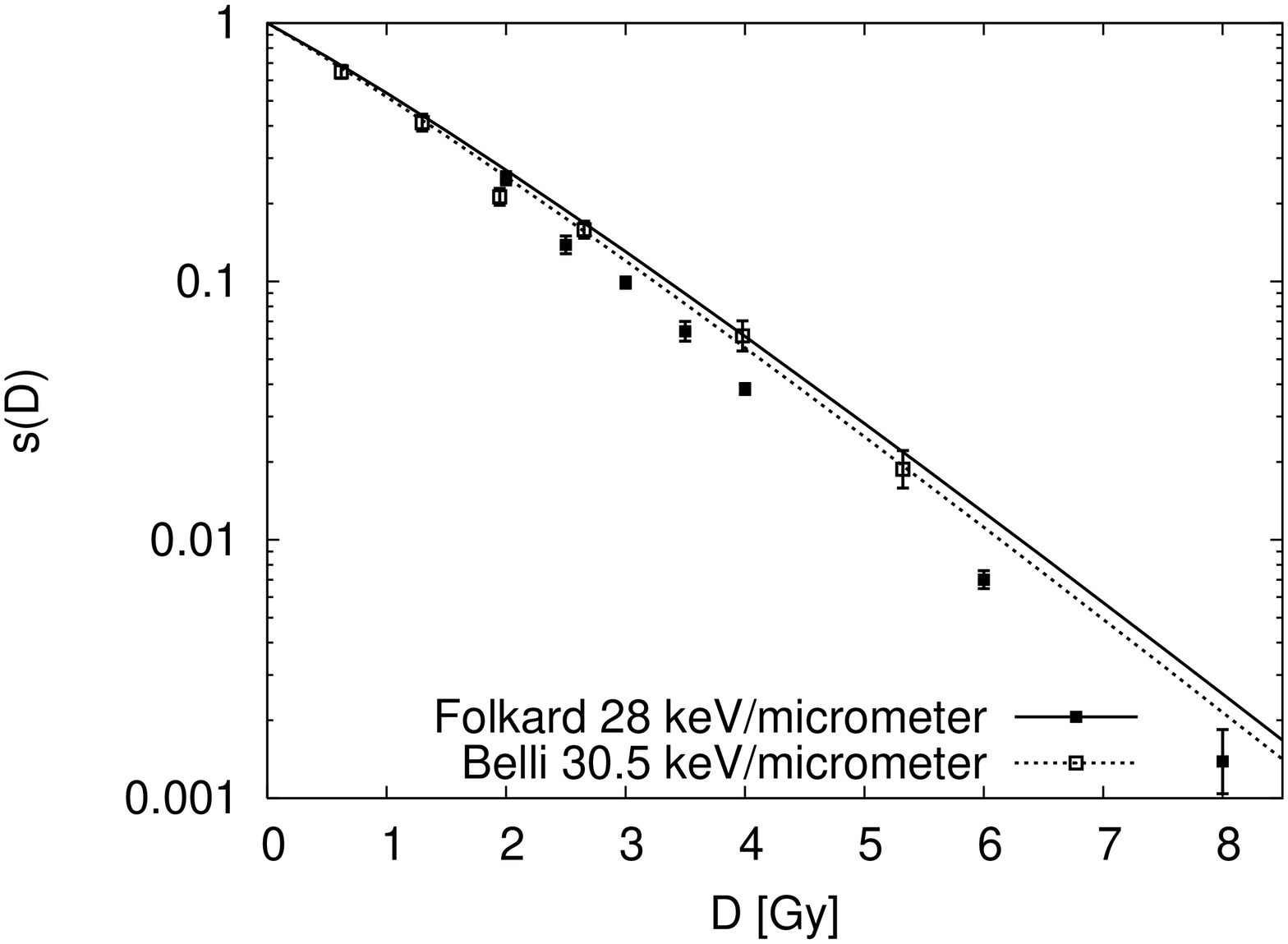}
		\includegraphics[trim=0cm 0.4cm 0cm 1.2cm, clip, angle=-90, width=0.32\textwidth]{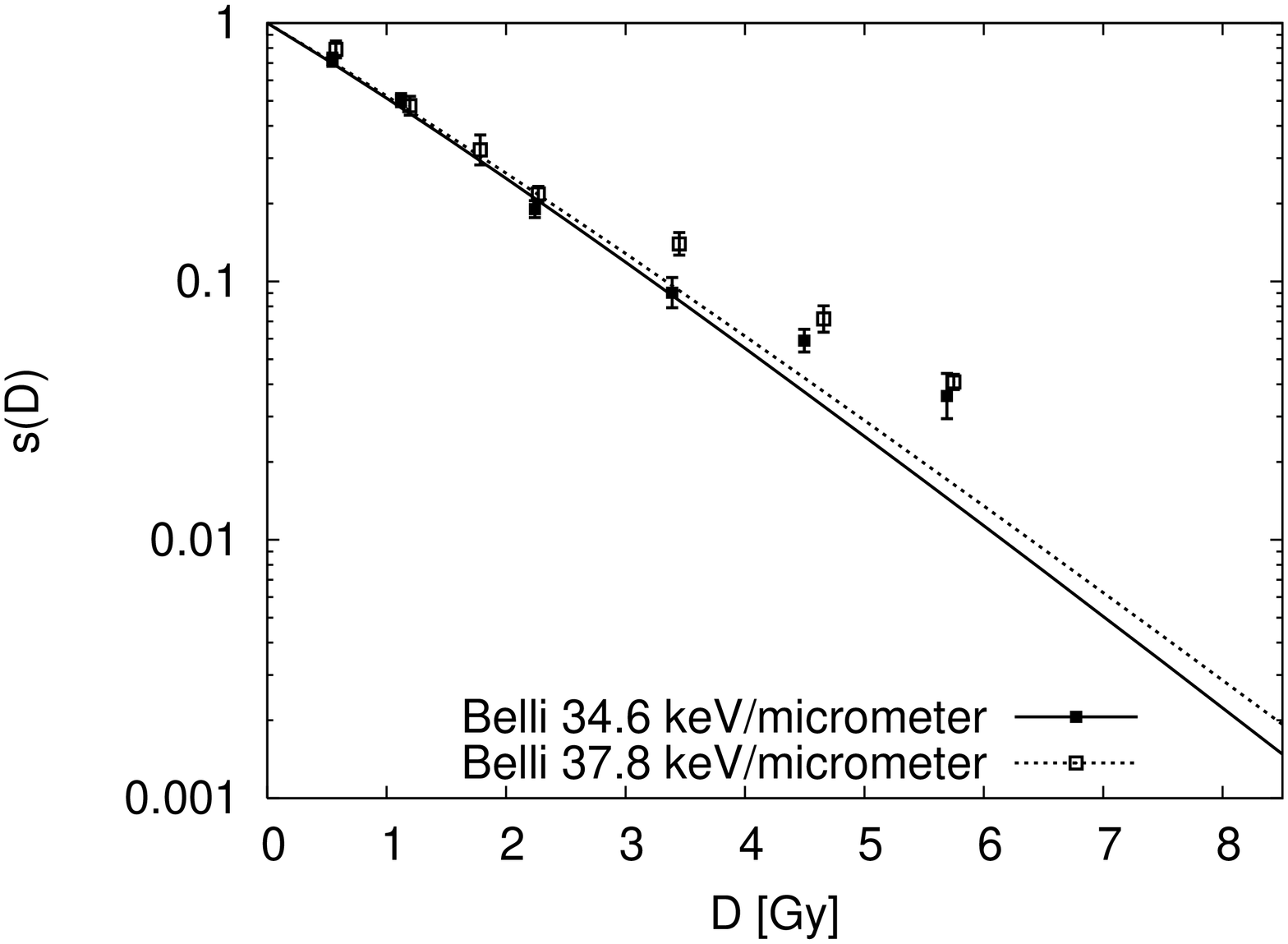}
	\caption{Survival curves for V79 cells irradiated by protons; data taken from \citep{Belli, Folkard96, Perris, Goodhead}.}
	\label{fig:p}
\end{figure}

\begin{figure}[!htb]
	\centering
		\includegraphics[trim=0cm 0.4cm 0cm 1.2cm, clip, angle=-90, width=0.32\textwidth]{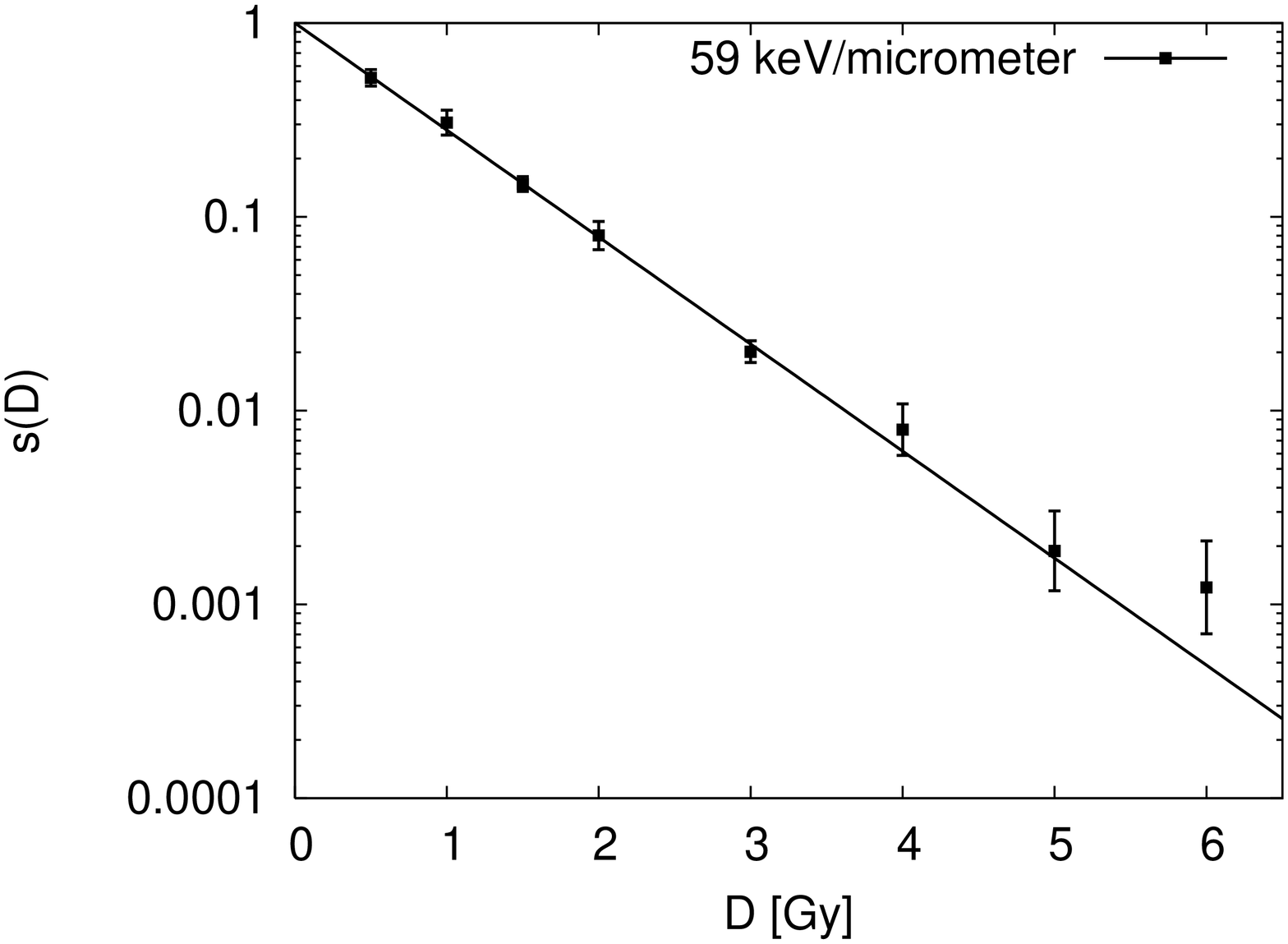}
		\includegraphics[trim=0cm 0.4cm 0cm 1.2cm, clip, angle=-90, width=0.32\textwidth]{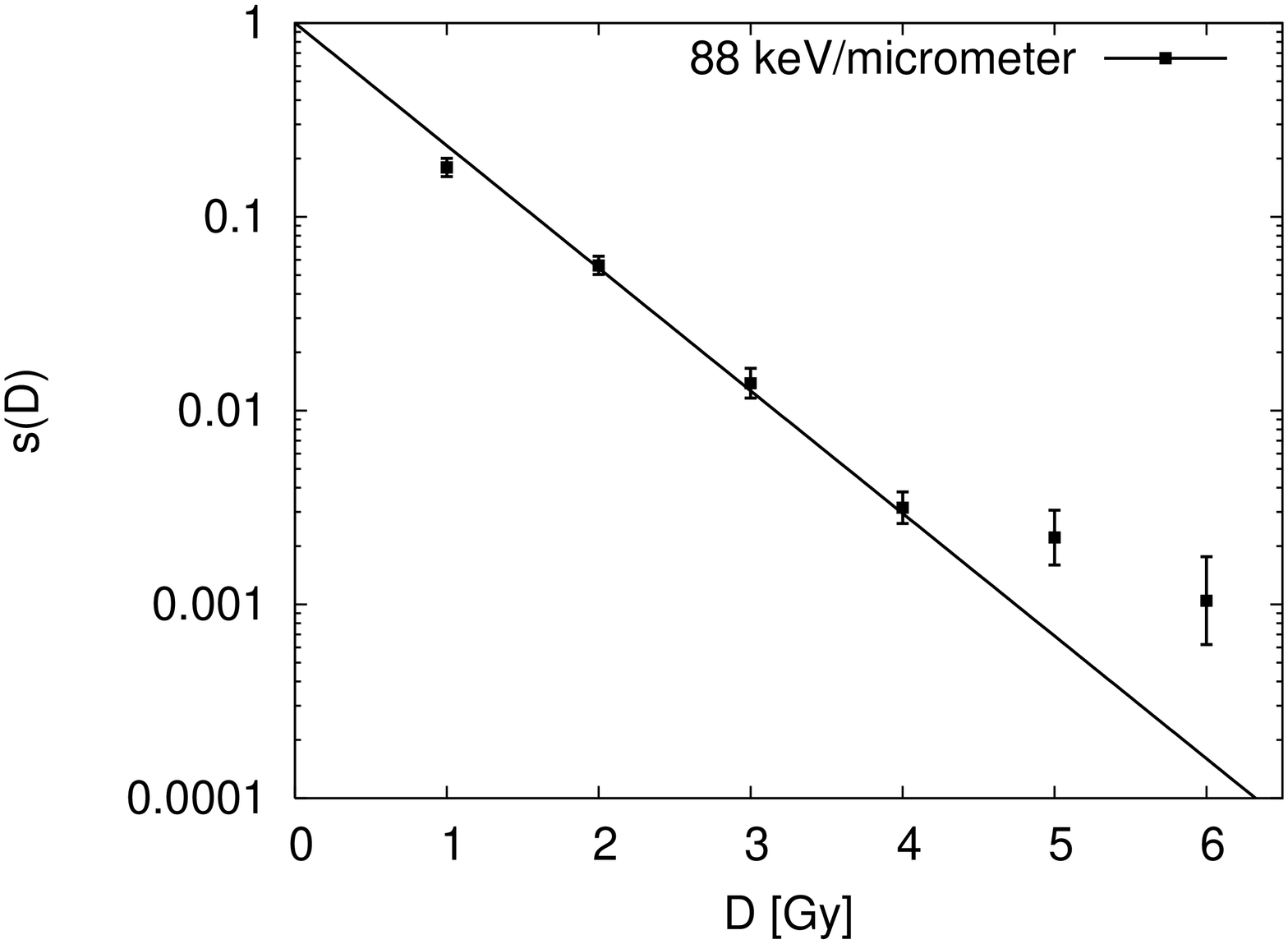}
		\includegraphics[trim=0cm 0.4cm 0cm 1.2cm, clip, angle=-90, width=0.32\textwidth]{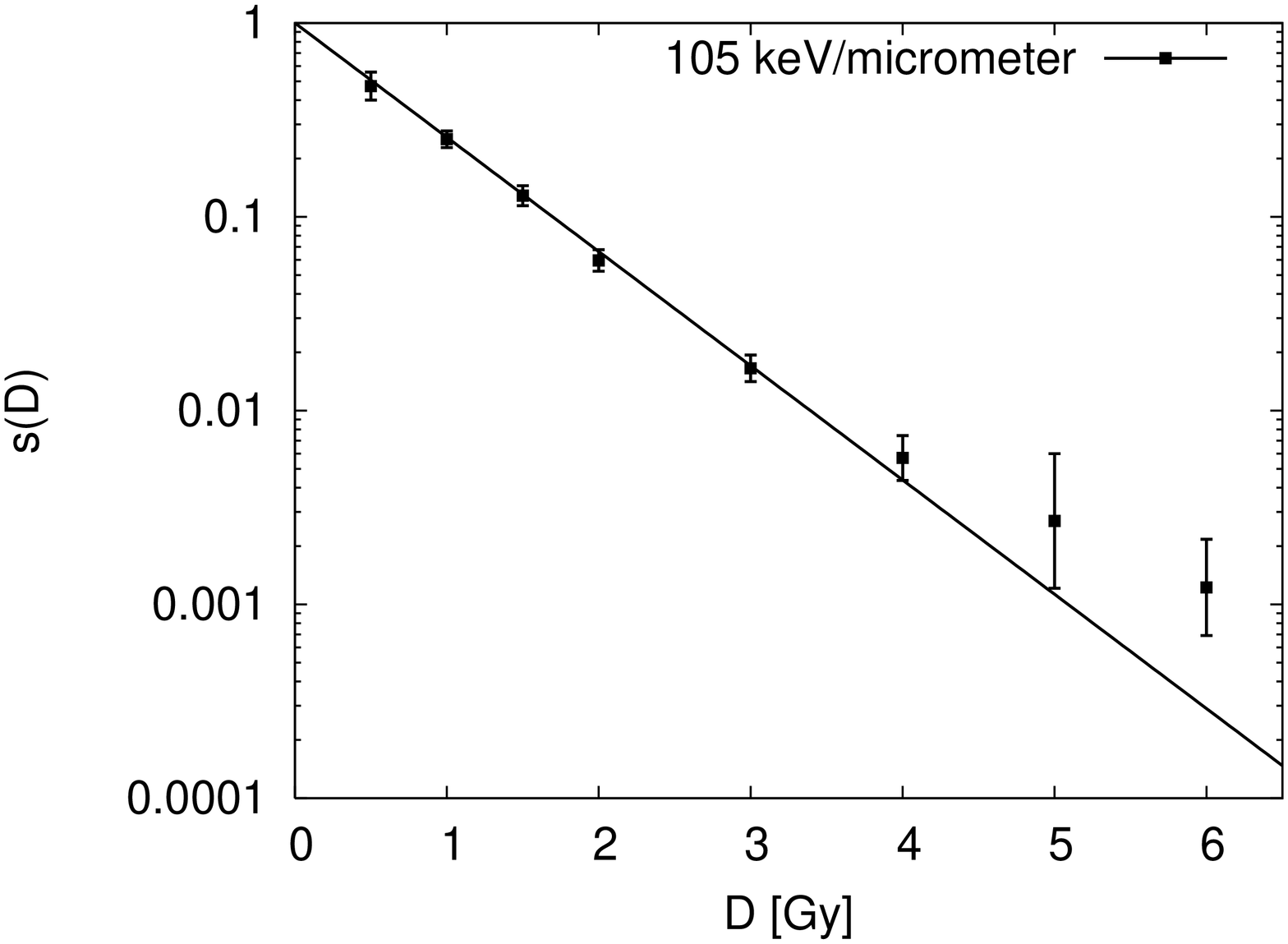}
	\caption{V79 cells irradiated by low-energy $\mathrm{^3He}$ ions; data from \citep{Folkard96}.}
	\label{fig:3He}
\end{figure}

\begin{figure}[!htb]
	\centering
		\includegraphics[trim=0cm 0.4cm 0cm 1.2cm, clip, angle=-90, width=0.32\textwidth]{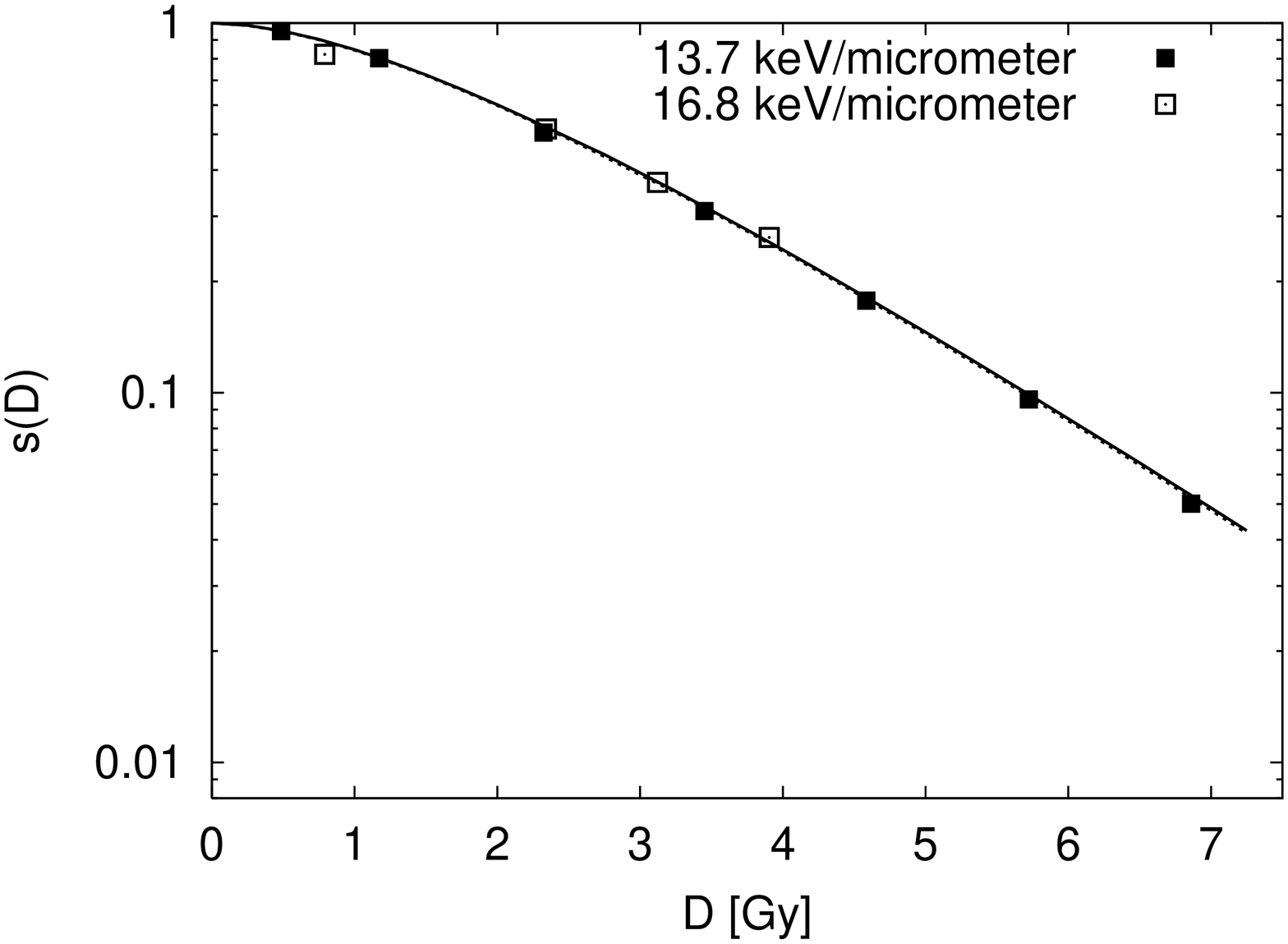}
		\includegraphics[trim=0cm 0.4cm 0cm 1.2cm, clip, angle=-90, width=0.32\textwidth]{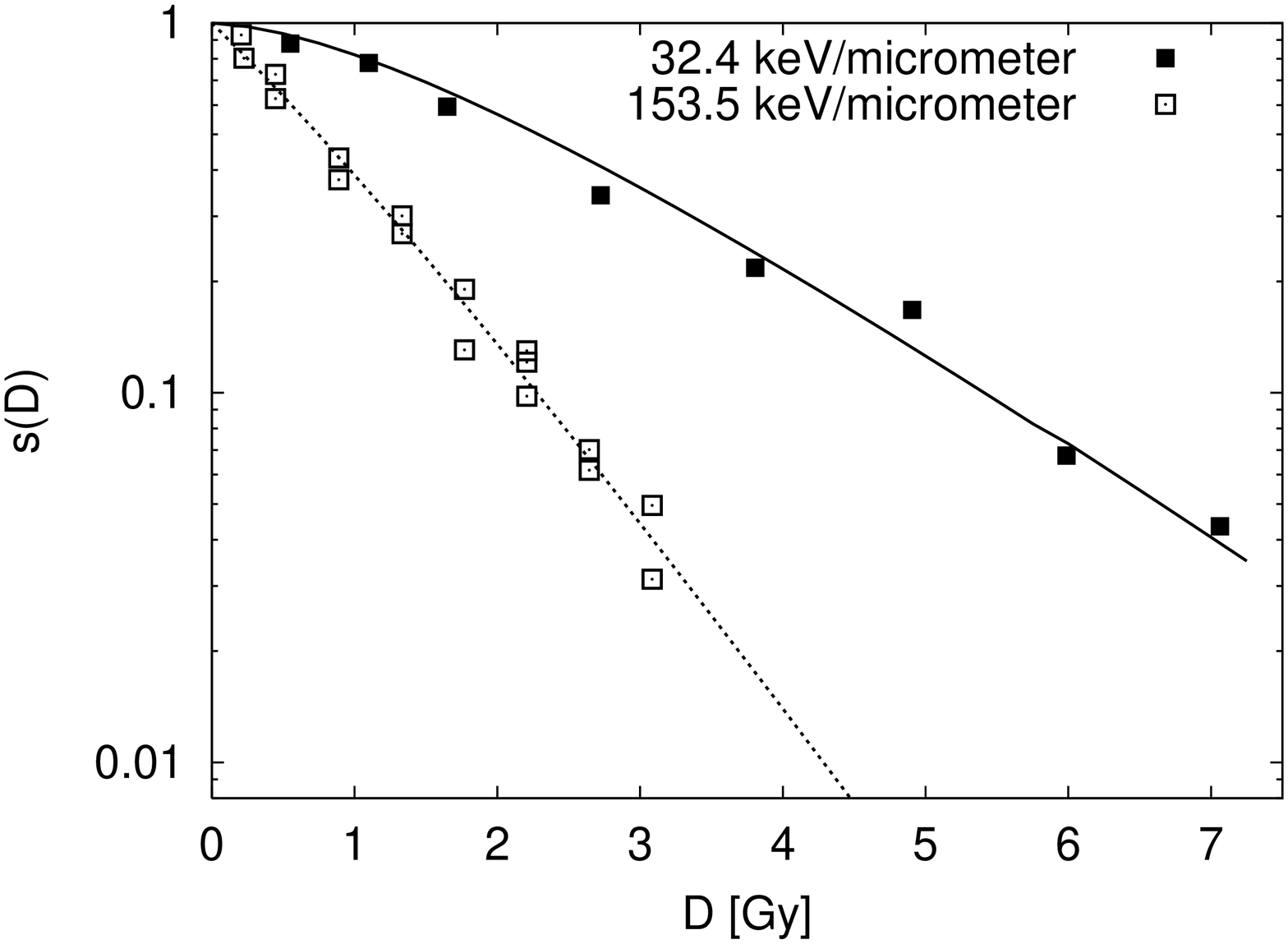}
		\includegraphics[trim=0cm 0.4cm 0cm 1.2cm, clip, angle=-90, width=0.32\textwidth]{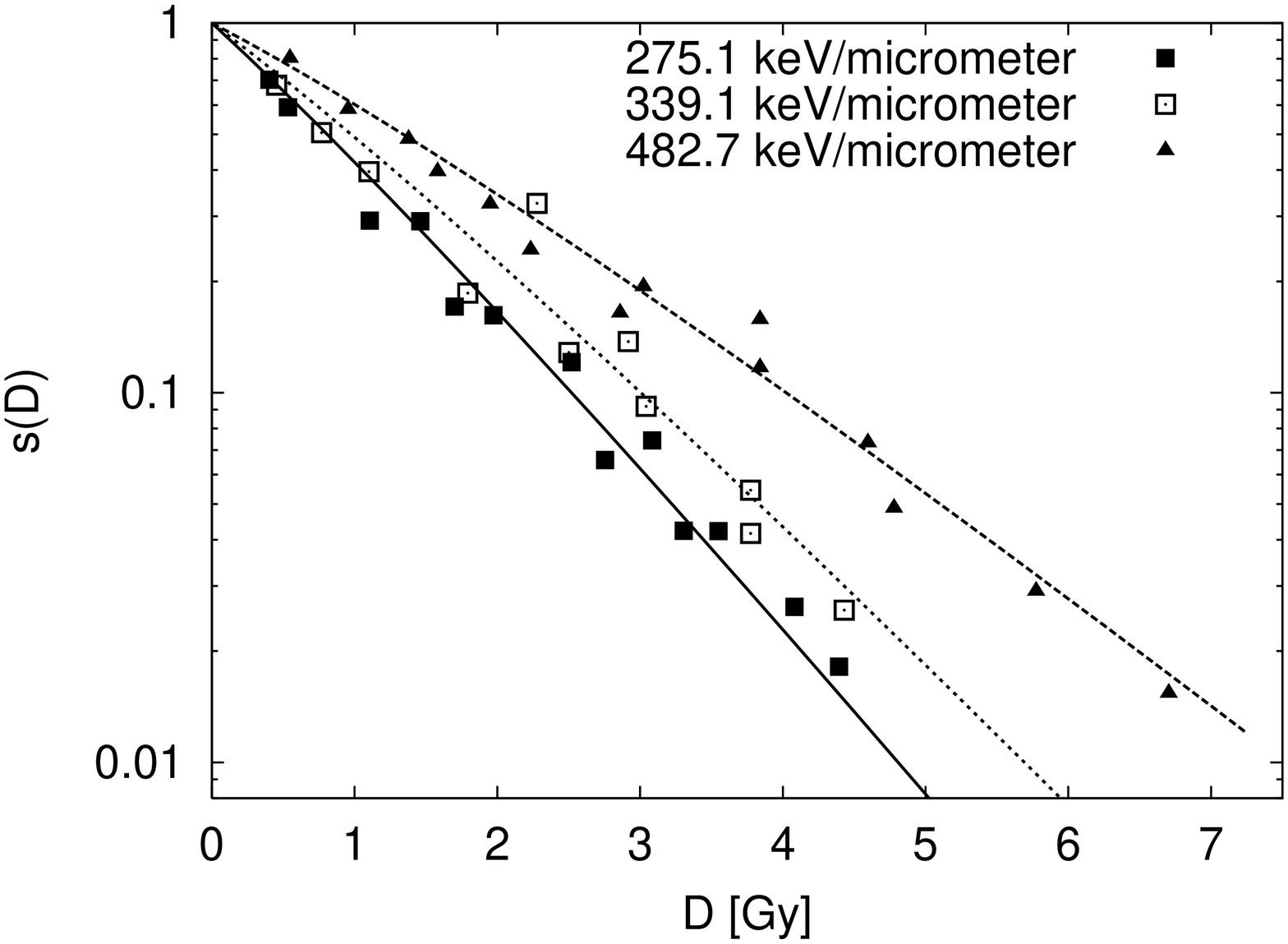}
	\caption{Representation of survival curves for V79 cells irradiated by carbon ions; data from \citep{Wilma}.}
	\label{fig:C}
\end{figure}

\begin{figure}[!htb]
	\centering
		\includegraphics[trim=0cm 0.4cm 0cm 1.2cm, clip, angle=-90, width=0.32\textwidth]{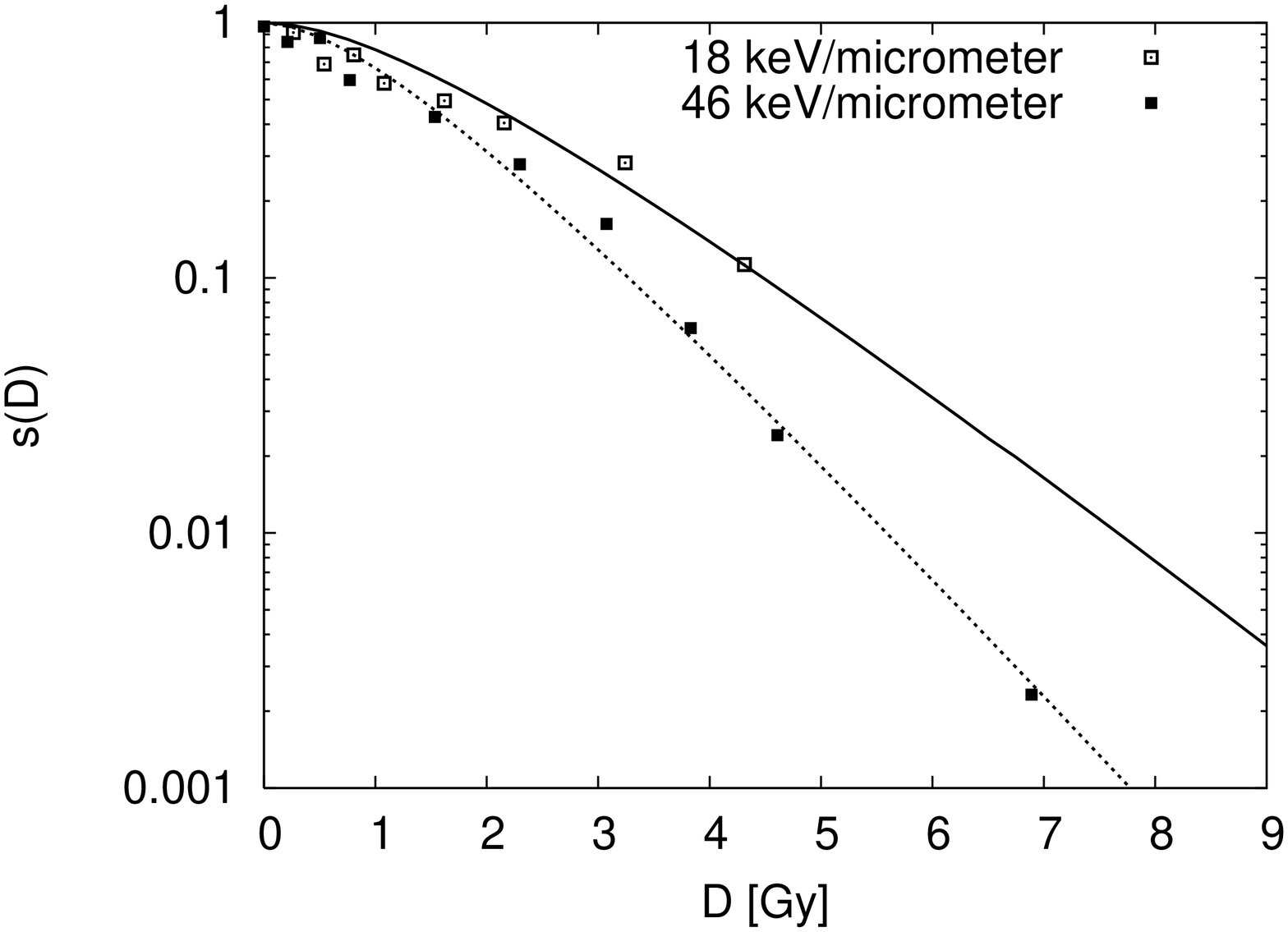}
		\includegraphics[trim=0cm 0.4cm 0cm 1.2cm, clip, angle=-90, width=0.32\textwidth]{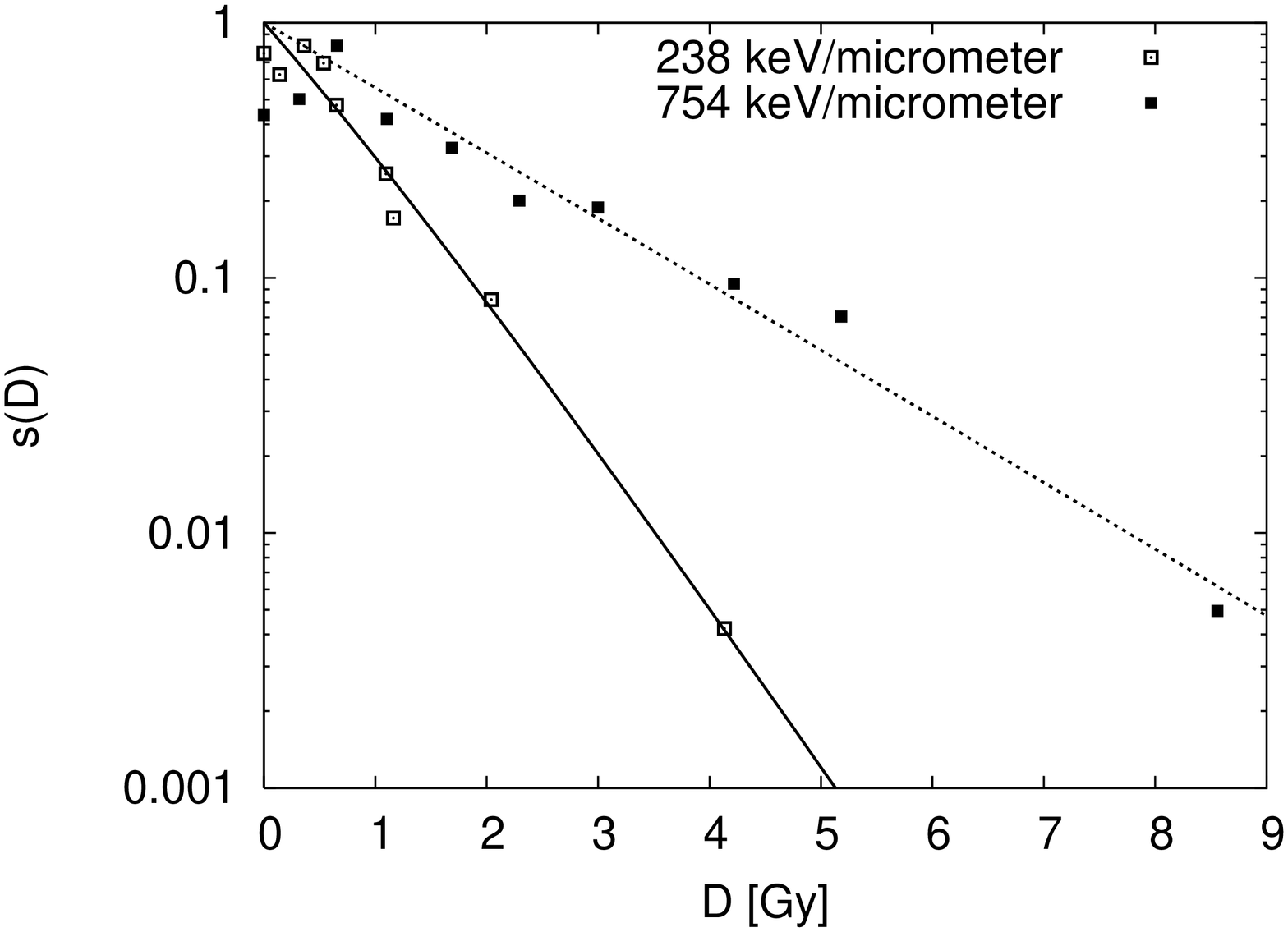}
	\caption{Survival of V79 cells irradiated by oxygen ions; data measured by \citet{Kiefer}.}
	\label{fig:O}
\end{figure}

\begin{figure}[!htb]
	\centering
		\includegraphics[angle=-90, width=0.34\textwidth]{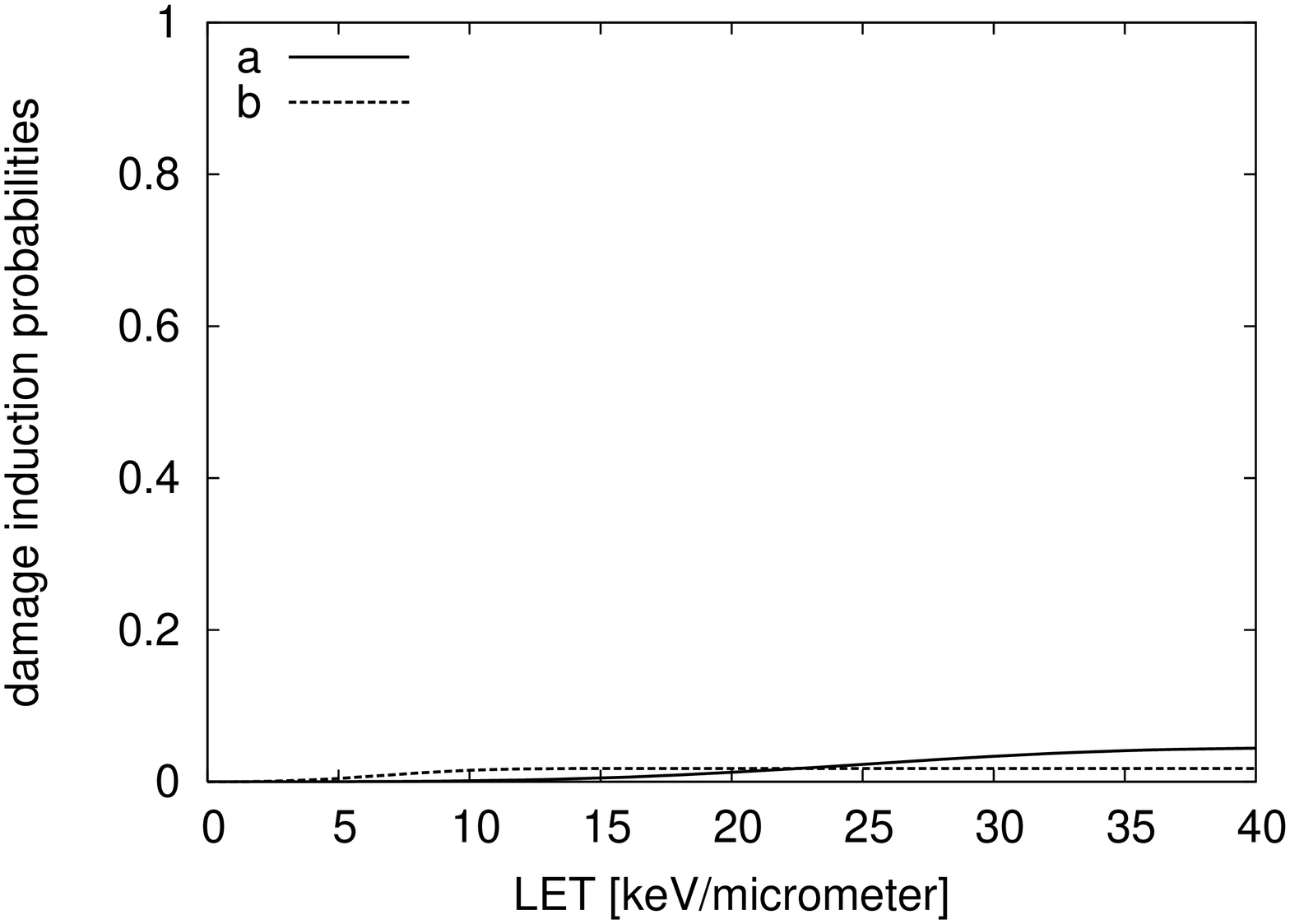}
		\includegraphics[angle=-90, width=0.34\textwidth]{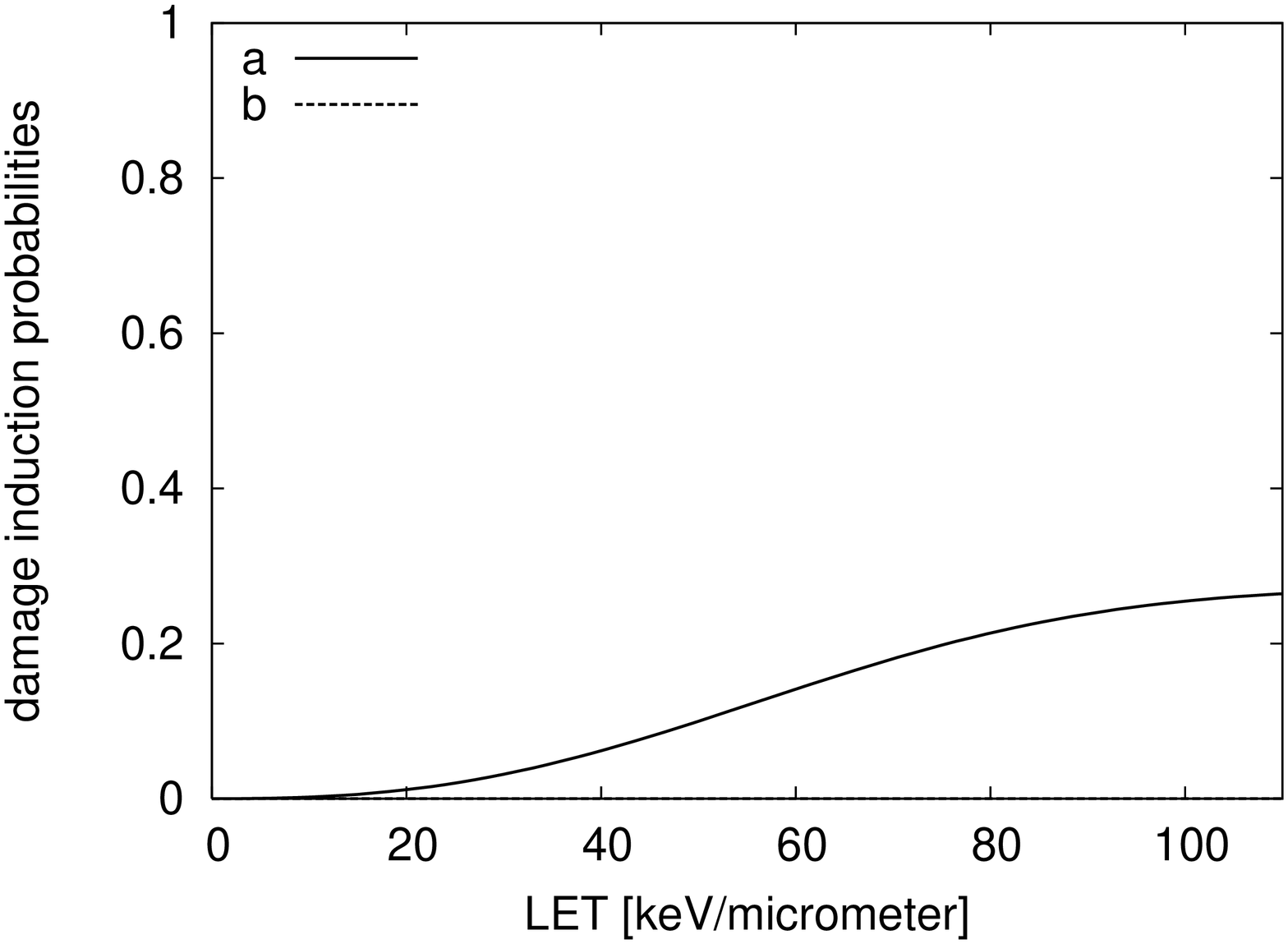}
		\includegraphics[angle=-90, width=0.34\textwidth]{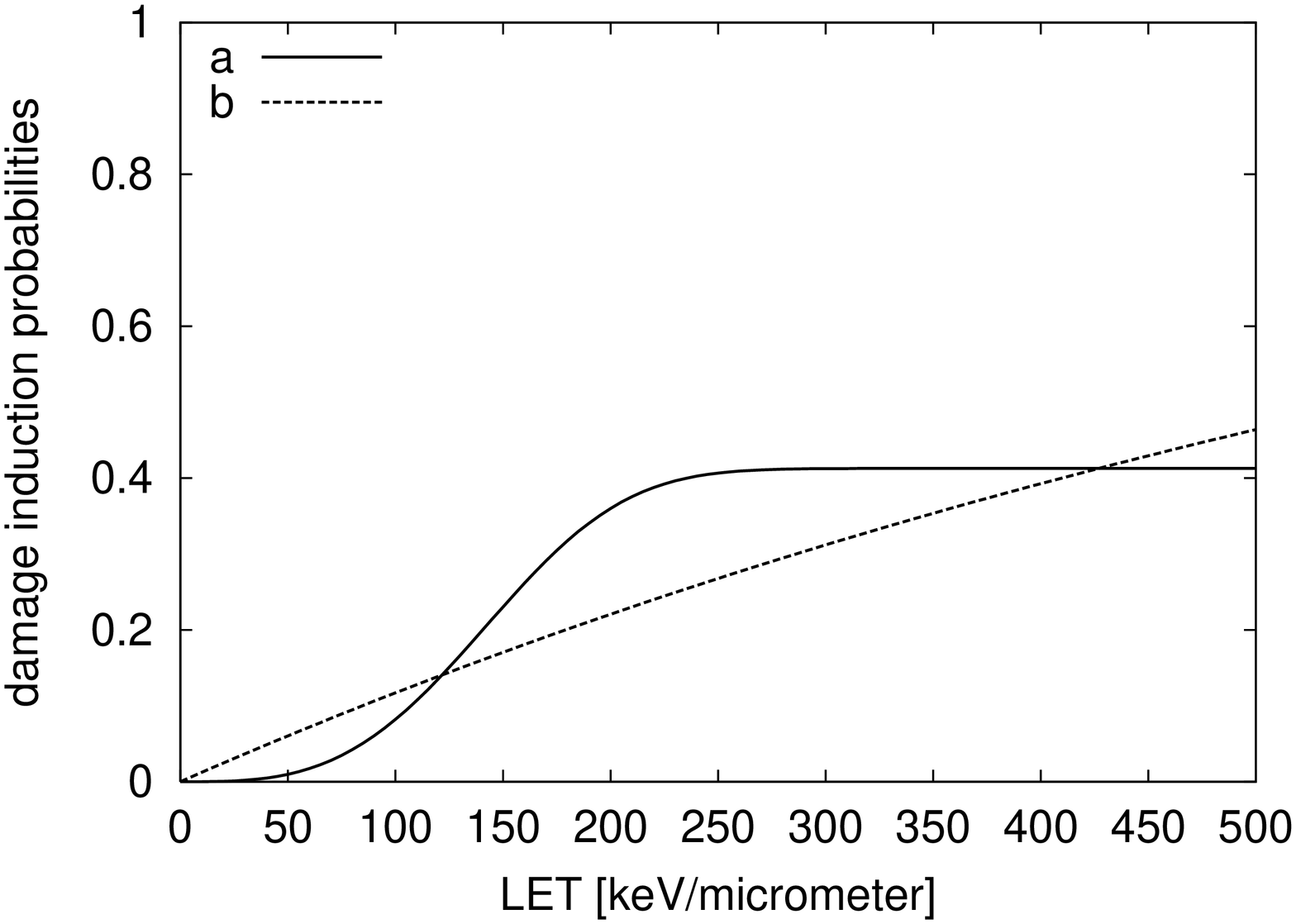}
		\includegraphics[angle=-90, width=0.34\textwidth]{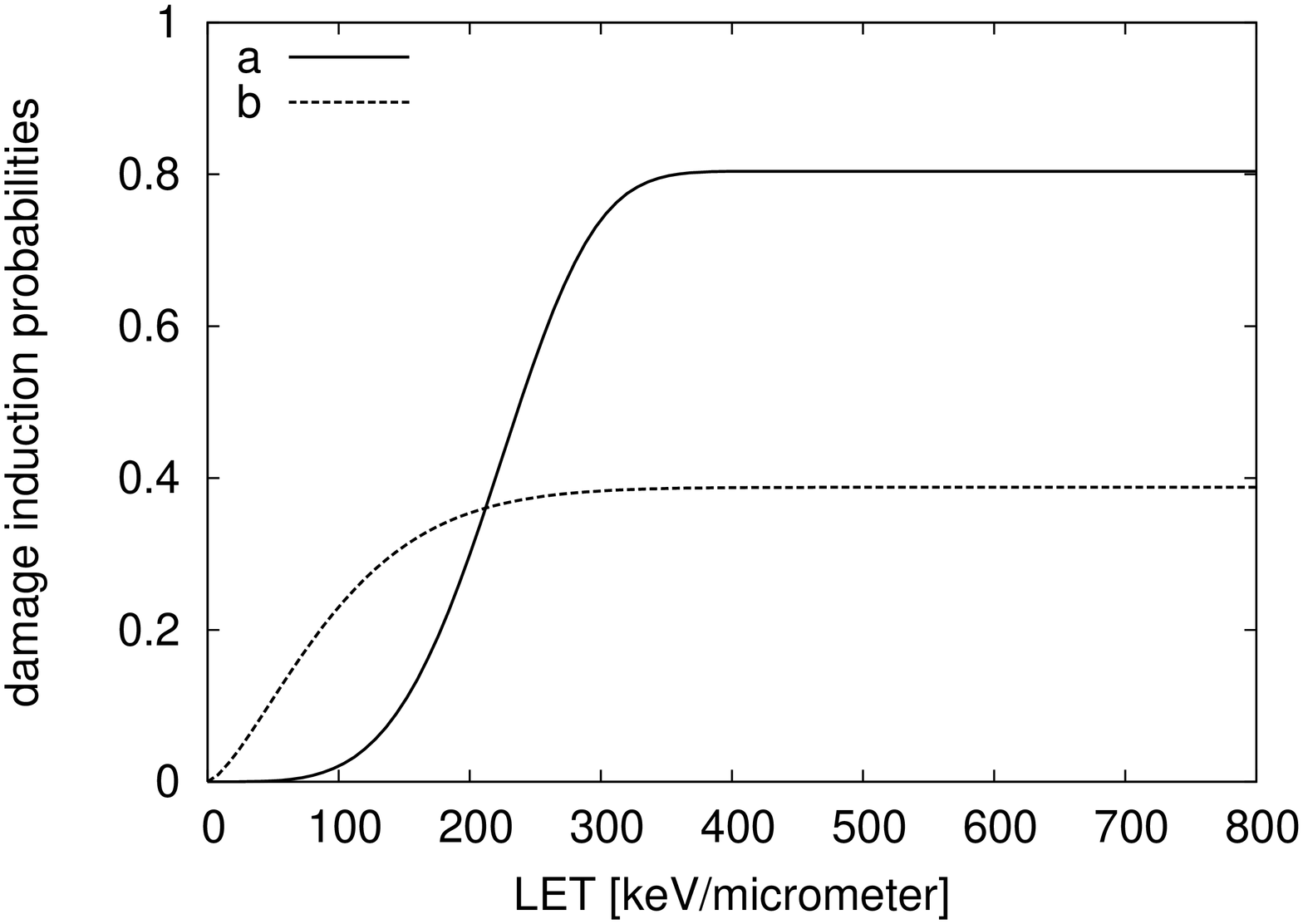}
	\caption{Probabilities of single-track and combined-damage induction for protons (top left), helium-3 (top right), carbon (bottom left) and oxygen ions (bottom right) in dependence on LET. For helium-3, the combined damages might have been neglected at all, $b=0$.}
	\label{fig:ab}
\end{figure}

Model representations of survival curves for V79 cells irradiated by different ions are shown in Figures~\ref{fig:p}-\ref{fig:O}. In Figure~\ref{fig:ab}, the probabilities of single-track and combined damage induction are plotted as functions of LET for different ions. Values of auxiliary model parameters are listed in Table~\ref{tab:ModelParameters}.

%%%%%%%%%%%%%%%%%%%%%%%%%%%%%%%%%%%%%%%%%%%%%%%%%%%%%%%%%%%%
%%%%%%%%%%%%%%%%%%%%%%%%%%%%%%%%%%%%%%%%%%%%%%%%%%%%%%%%%%%%

%%%%%%%%%%%%%%%%%%%%%%%%%%%%%%%%%%%%%%%%%%%%%%%%%%%%%%%%%%%%
%%%%%%%%%%%%%%%%%%%%%%%%%%%%%%%%%%%%%%%%%%%%%%%%%%%%%%%%%%%%

%%%%%%%%%%%%%%%%%%%%%%%%%%%%%%%%%%%%%%%%%%%%%%%%%%%%%%%%%%%%
%%%%%%%%%%%%%%%%%%%%%%%%%%%%%%%%%%%%%%%%%%%%%%%%%%%%%%%%%%%%

%%%%%%%%%%%%%%%%%%%%%%%%%%%%%%%%%%%%%%%%%%%%%%%%%%%%%%%%%%%%
%%%%%%%%%%%%%%%%%%%%%%%%%%%%%%%%%%%%%%%%%%%%%%%%%%%%%%%%%%%%

%%%%%%%%%%%%%%%%%%%%%%%%%%%%%%%%%%%%%%%%%%%%%%%%%%%%%%%%%%%%
%%%%%%%%%%%%%%%%%%%%%%%%%%%%%%%%%%%%%%%%%%%%%%%%%%%%%%%%%%%%

\begin{table}
\caption{\label{tab:ModelParameters}Values of model parameters, describing the dependence of damage induction probabilities on ion LET values for V79 and CHO-K1 cells, Eq.\ (\ref{ab}).}
\begin{indented}
\item[]\begin{tabular}{@{}l|llll|l}
\br
&&&V79&&CHO-K1\\
\mr
&p&$^3$He&C&O&C\\
\mr
$a_0$&0.045&0.25&0.41&0.80&0.50\\
$a_1$ [$\mu$m/keV]&0.036&0.015&0.0063&0.0042&0.0064\\
$a_2$&3.57&2.79&3.21&4.17&2.0\\
$b_0$&0.017&0.0&1.0&0.39&0.26\\
$b_1$ [$\mu$m/keV]&0.13&&0.0012&0.0093&0.0011\\
$b_2$&2.81&&1.0&1.44&0.69\\
\br
\end{tabular}
\end{indented}
\end{table}

%%%%%%%%%%%%%%%%%%%%%%%%%%%%%%%%%%%%%%%%%%%%%%%%%%%%%%%%%%%%

\begin{figure}[!htb]
	\centering
		\includegraphics[trim=0cm 0.4cm 0cm 1.2cm, clip, angle=-90, width=0.34\textwidth]{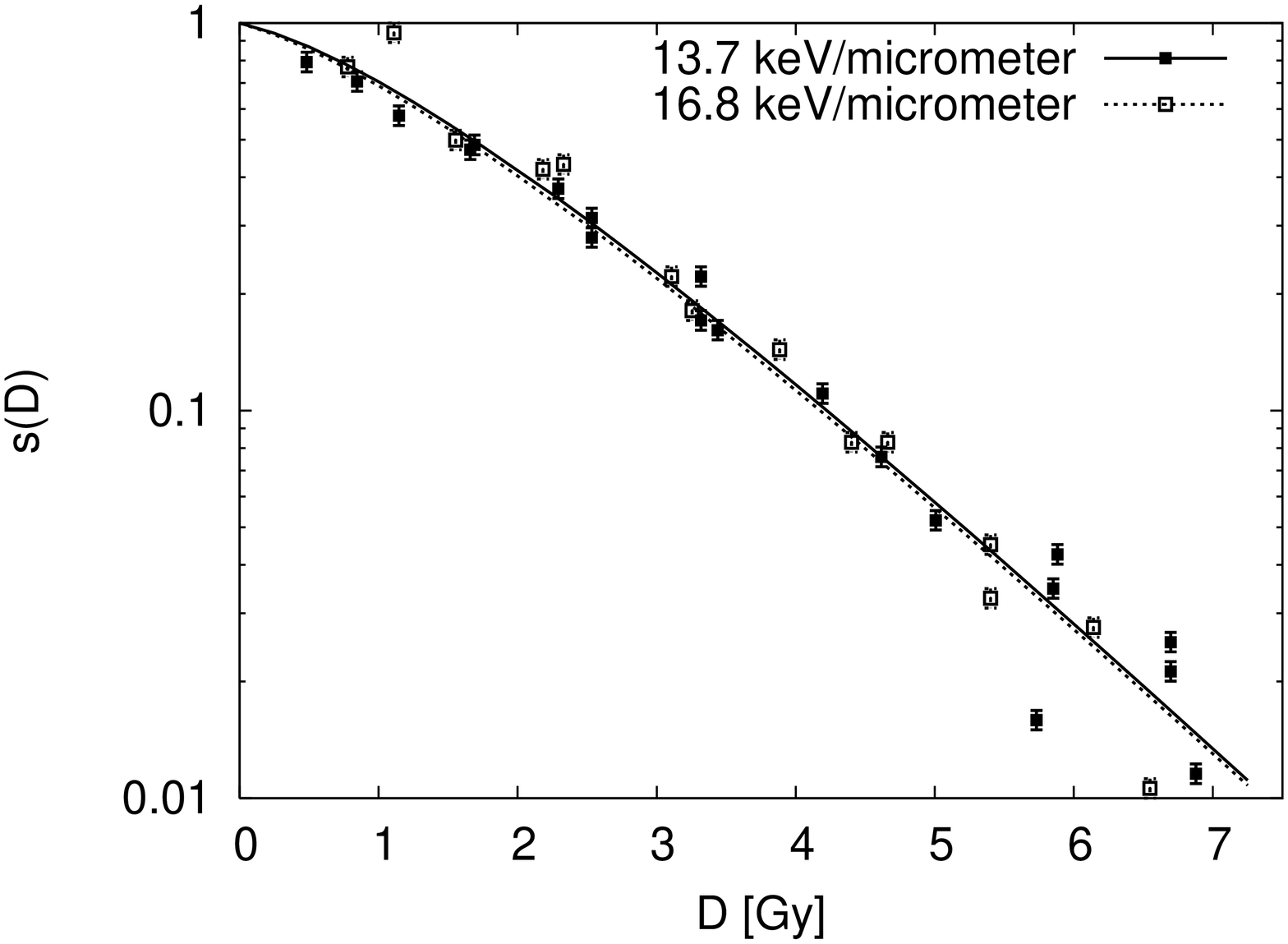}
		\includegraphics[trim=0cm 0.4cm 0cm 1.2cm, clip, angle=-90, width=0.34\textwidth]{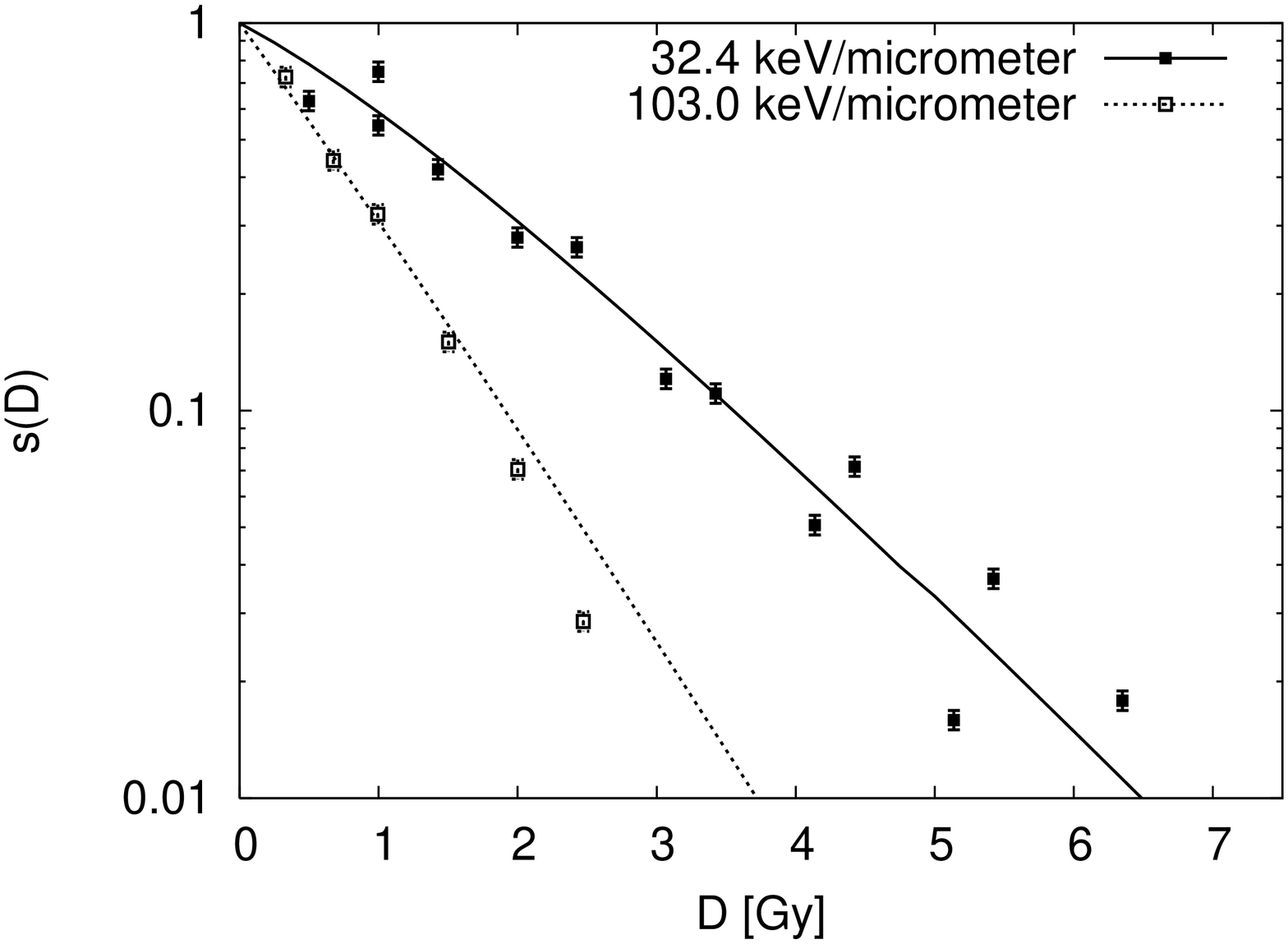}
		\includegraphics[trim=0cm 0.4cm 0cm 1.2cm, clip, angle=-90, width=0.34\textwidth]{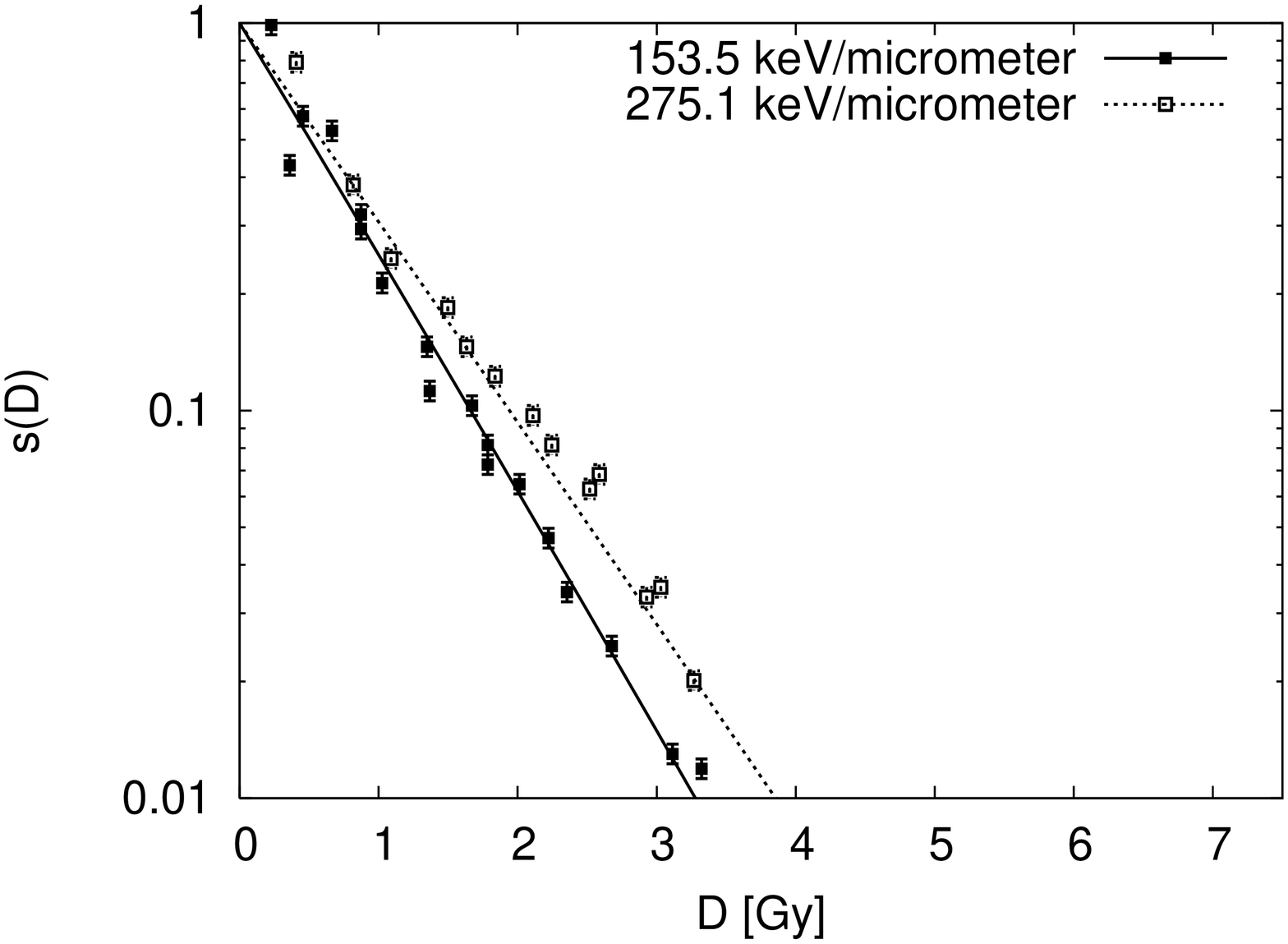}
		\includegraphics[trim=0cm 0.4cm 0cm 1.2cm, clip, angle=-90, width=0.34\textwidth]{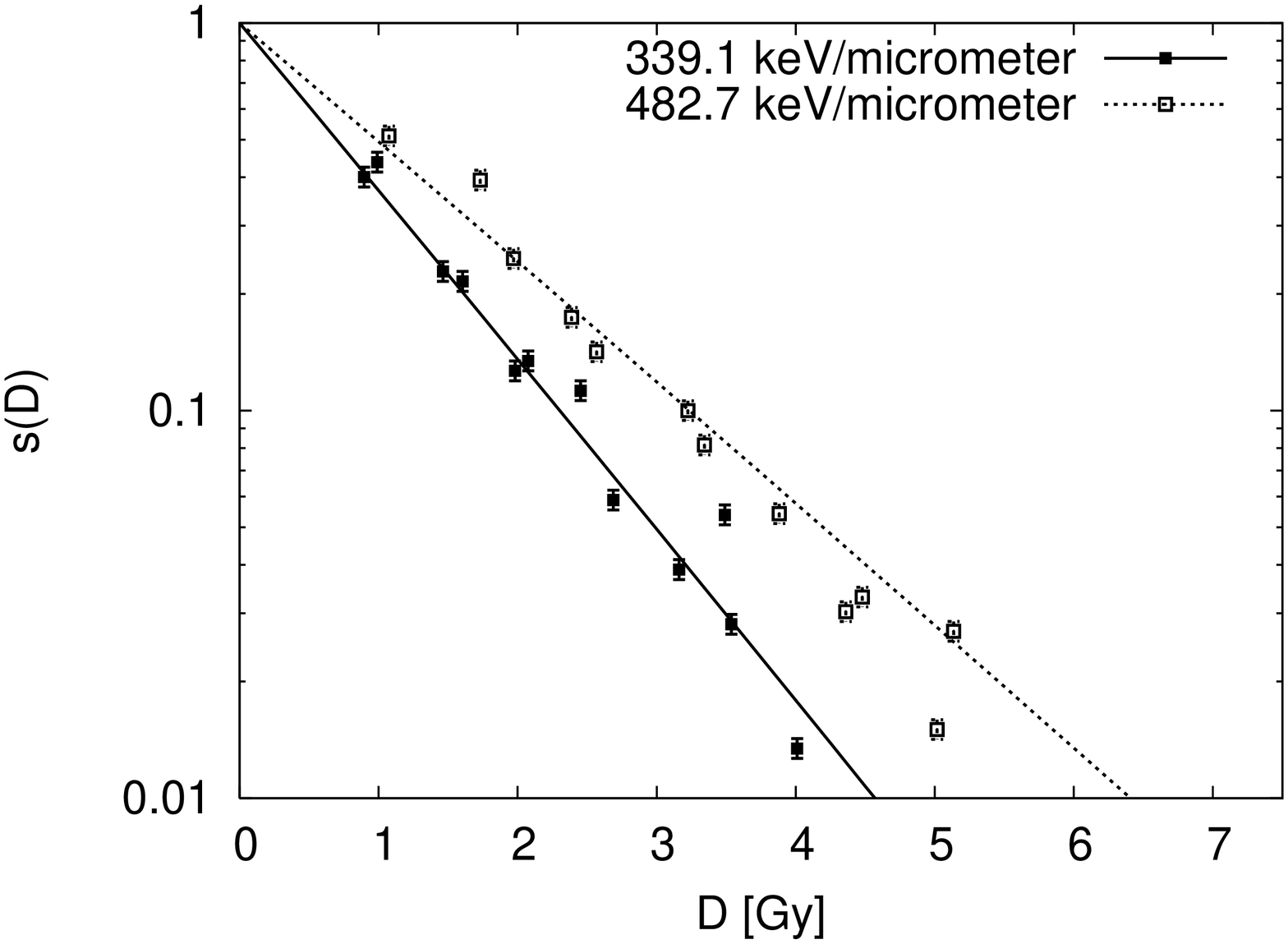}
		\includegraphics[angle=-90, width=0.34\textwidth]{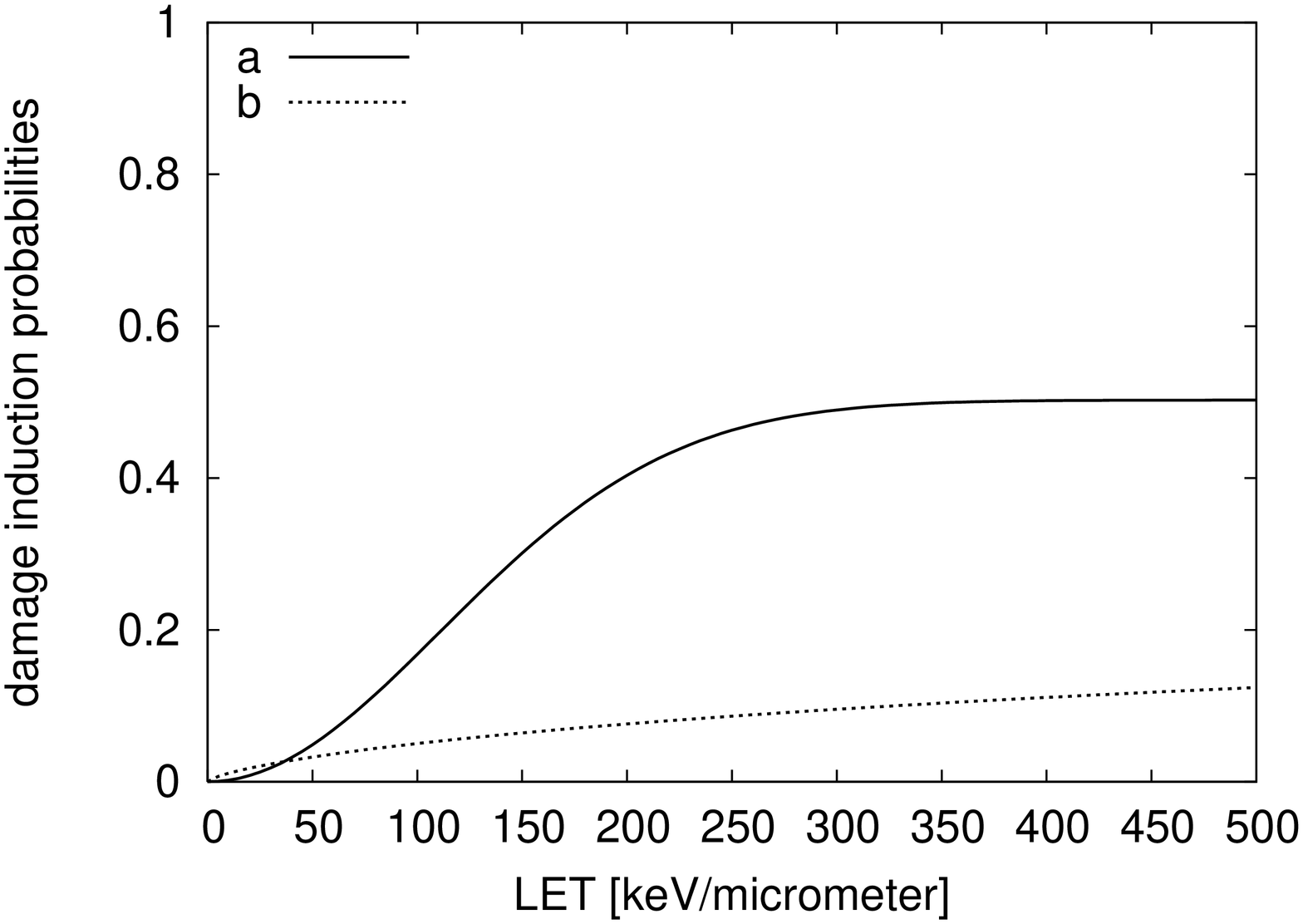}
	\caption{Survival curves (data from \citep{Wilma}) and damage induction probabilities for CHO-K1 cells irradiated by carbon ions.}
	\label{fig:C-CHO}
\end{figure}

A similar analysis has been performed for survival data of CHO-K1 cells irradiated by carbon ions, measured experimentally by \citet{Wilma}. Model representation of survival curves and the derived damage induction probabilities are shown in Figure~\ref{fig:C-CHO}; the parameter values are given in Table~\ref{tab:ModelParameters}.

%%%%%%%%%%%%%%%%%%%%%%%%%%%
%%%%%%%%%%%%%%%%%%%%%%%%%%%

The present results correspond to the geometrical cross section for V79 cells chosen uniquely for all data sets to be $\sigma$ = 87.8 $\mathrm{\mu m^2}$, based on the value reported in \citep{Wilma}, in order to enable comparisons between different ions. Values in the range of 50~--~135 $\mathrm{\mu m^2}$, also reported in the literature (compare e.g. \citep{Kraft-Scholz-50, Belli}), have yielded similar results (figures not shown here). For CHO cells, the value of geometrical cross section $\sigma$ = 108 $\mathrm{\mu m^2}$ was taken, as reported in the given experiment \citep{Wilma}.

%%%%%%%%%%%%%%%%%%%%%%%%%%%%%%%%%%%%%%%%%%%%%%%%%%%%%%%%%%%%%%%%%%%%%%%%%%%%%%%%
\section{Discussion}
\label{sec:Discussion}

%%%%%%%%%%%%%%%%%%%%%%%%%%%%%%%%%%%%%%%%%%%%%%%%%%%%%%%%%%%%%%%%%%%%%%%%%%%%%%%%
\subsection{Single-track damage induced by protons: Possible indications for differences in radiobiological mechanisms between protons and ions}
\label{sec:DiscussionProtons}

\begin{figure}[!htb]
	\centering
		\includegraphics[trim=0cm 0.4cm 0cm 1.2cm, clip, angle=-90, width=0.32\textwidth]{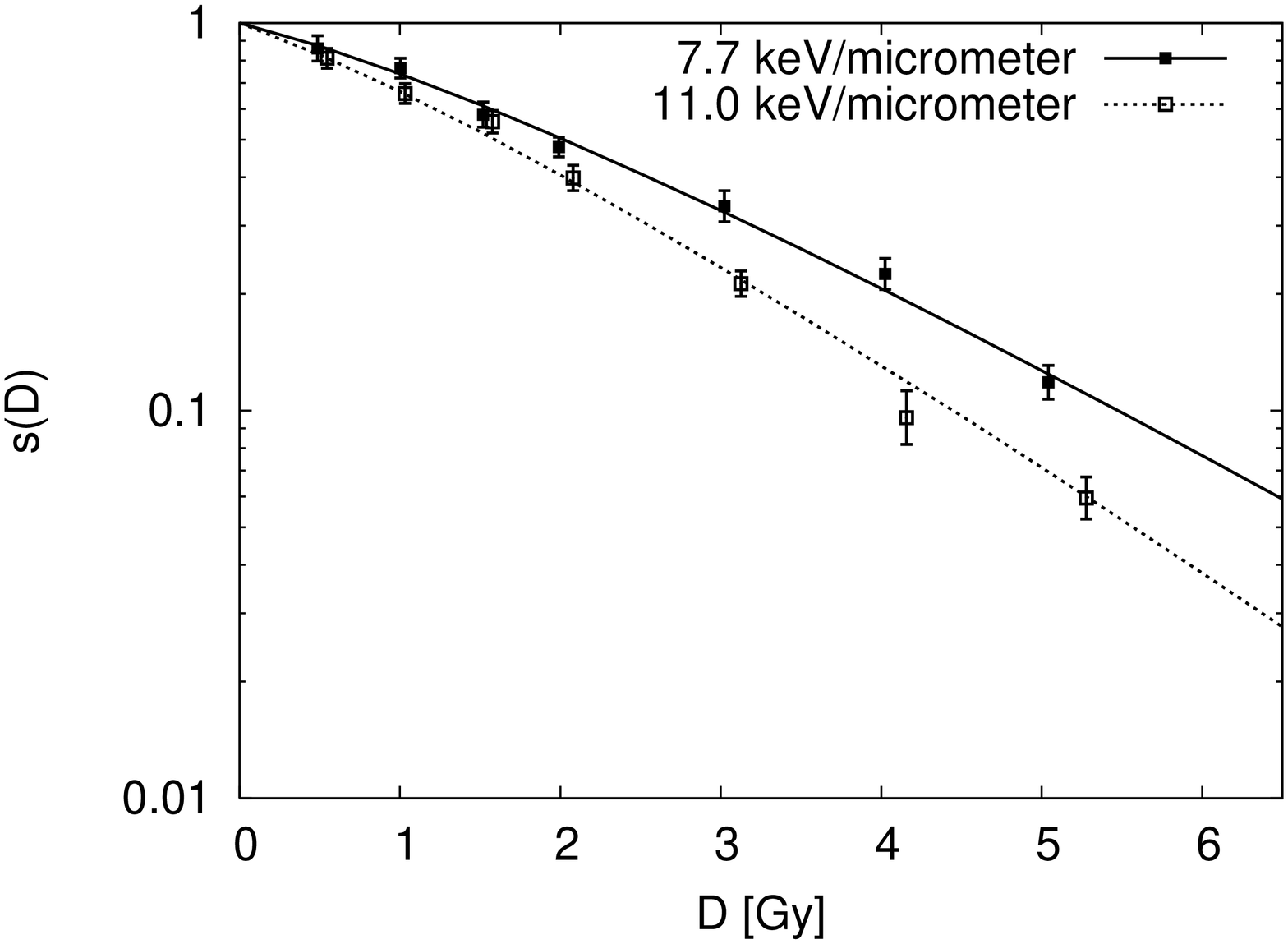}
		\includegraphics[trim=0cm 0.4cm 0cm 1.2cm, clip, angle=-90, width=0.32\textwidth]{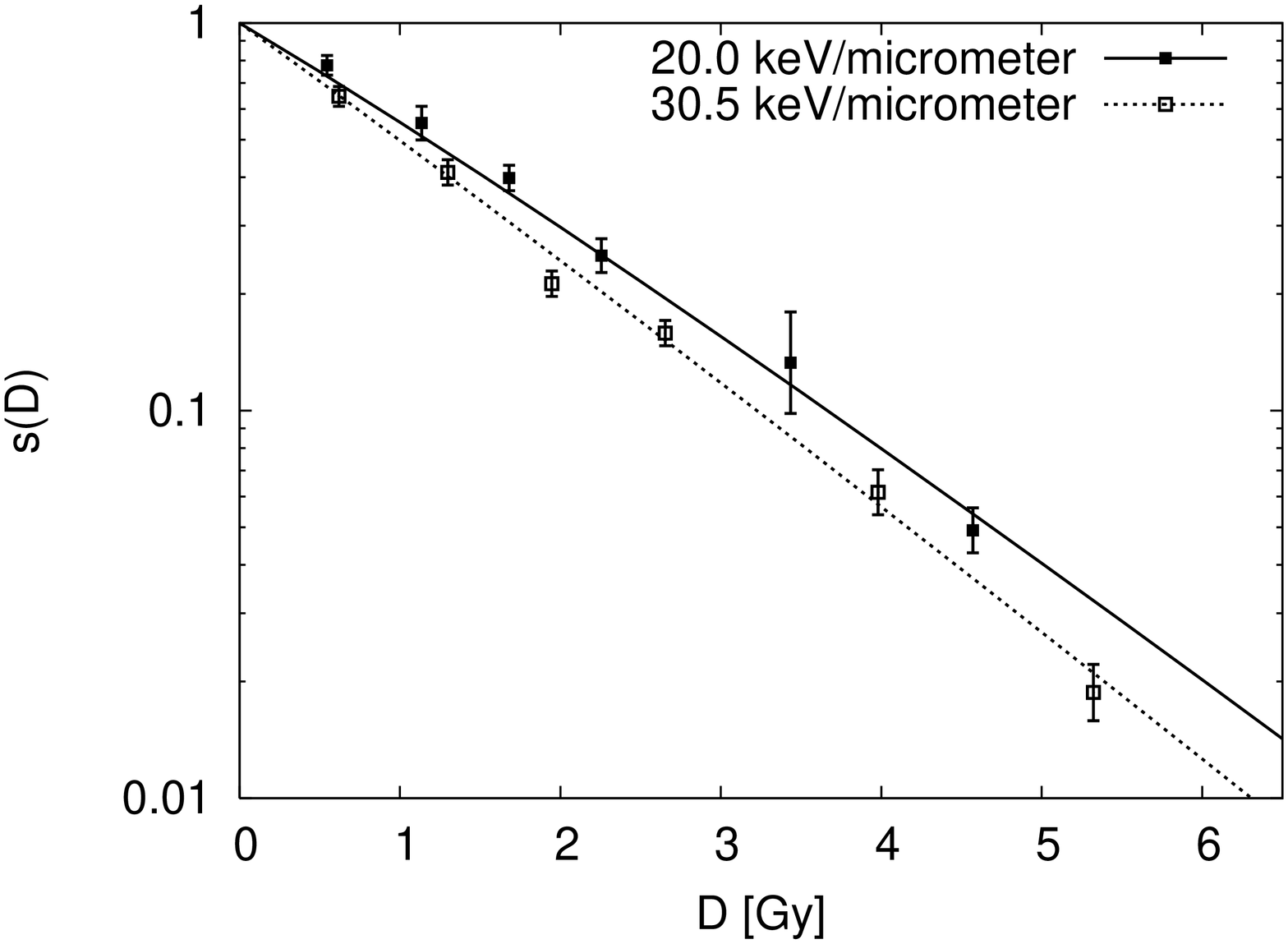}
		\includegraphics[trim=0cm 0.4cm 0cm 1.2cm, clip, angle=-90, width=0.32\textwidth]{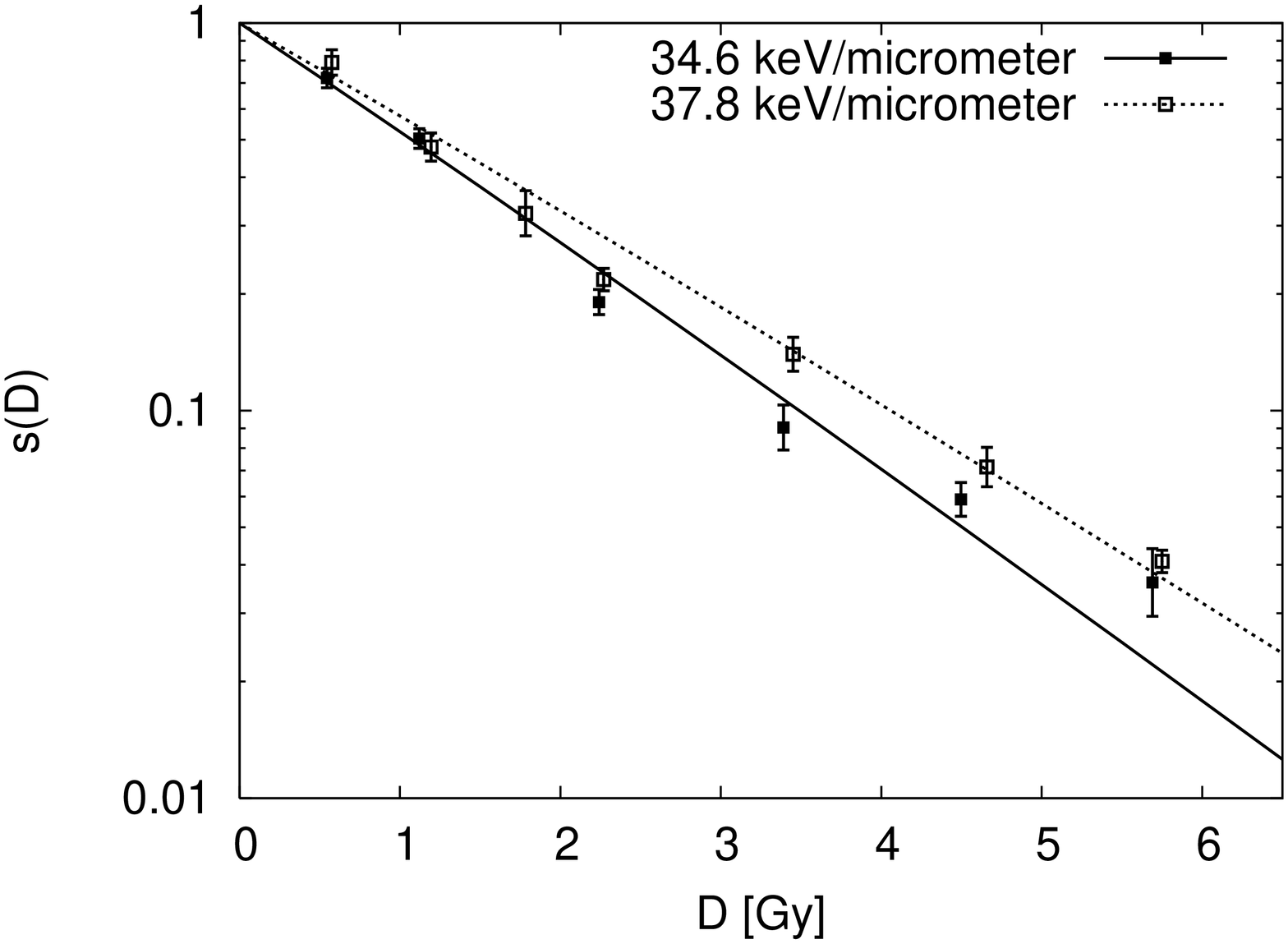}
		\includegraphics[angle=-90, width=0.32\textwidth]{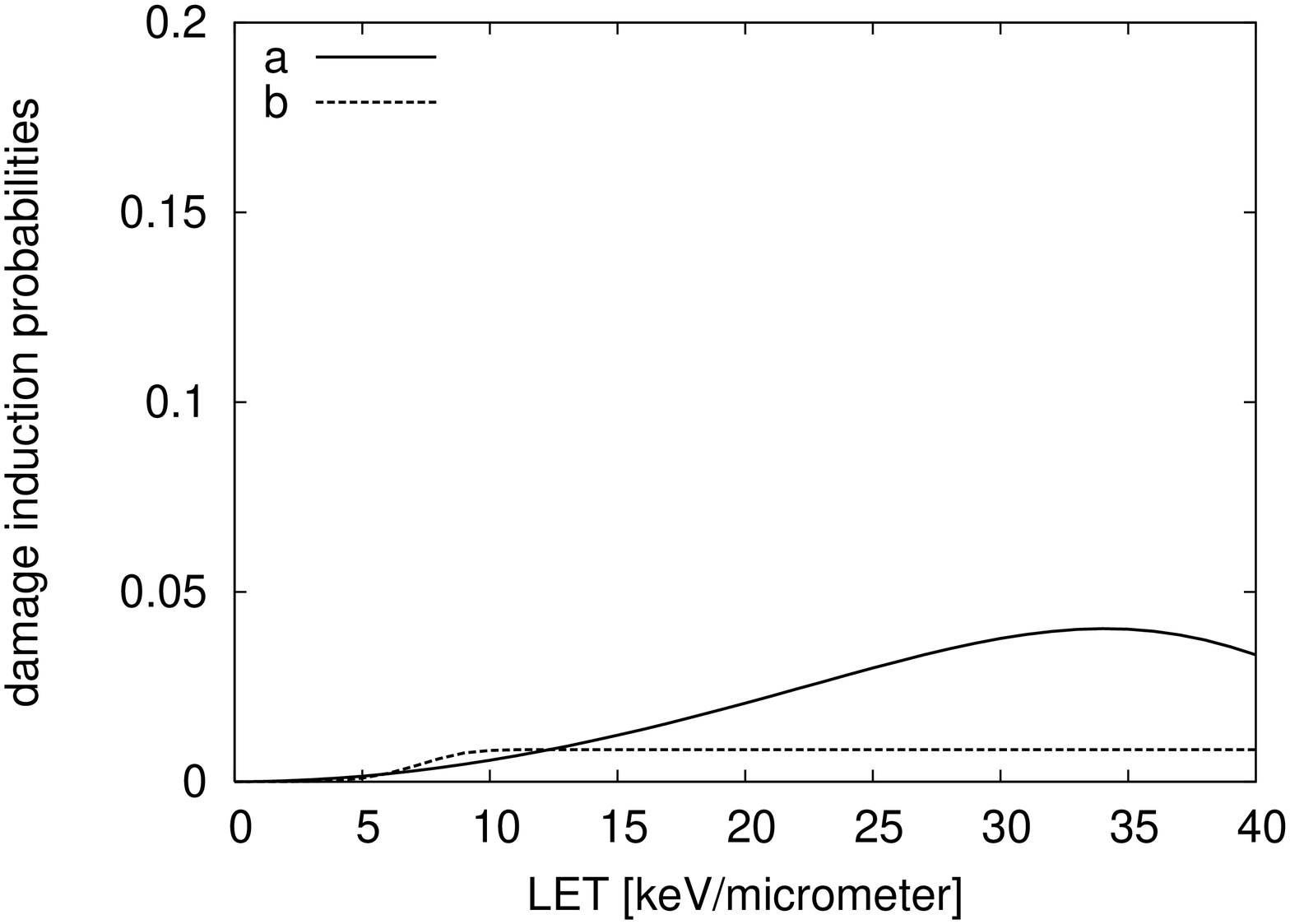}
	\caption{Detailed fits to proton-induced V79 inactivation, indicating a decrease in the single-track damage formation for protons at LET values above approximately 30 keV$/\mu$m; data measured by \citet{Belli}, calculations based on Eq.\ (\ref{ab-nonmonotonous}).}
	\label{fig:p-nonmonotonous}
\end{figure}

Measured data for protons of LET values above approximately 30 keV$/\mu$m \citep{Belli} suggest that the single-track damage formation probability, $a$, might decrease in this region; compare Figure~\ref{fig:p-nonmonotonous} where more detailed fits to the data are shown, based on the parameterization
\begin{eqnarray} \label{ab-nonmonotonous}
	a(\lambda) &=& a_0 (1 - \exp(-(a_1 \lambda)^{a_2})) \exp(-(a_3 \lambda)^{a_4}) \ , \nonumber \\ 
	b(\lambda) &=& b_0 (1 - \exp(-(b_1 \lambda)^{b_2})) \ .
\end{eqnarray}
Whereas the value of the goodness-of-fits criterion is $\chi^2=350$ for the data of \citet{Belli} in the joint fits to proton-induced survival shown in Figure~\ref{fig:p}, the corresponding value for the detailed fits shown in Figure~\ref{fig:p-nonmonotonous} is $\chi^2_{\mathrm{detailed}}=55$. The goodness-of-fits to this data set can be, in fact, further improved if, in addition to a decreasing single-track damage probability, the combined-damages are allowed to dominate in the whole protons' LET range and their repair is taken into account, compare \citep{PhysMedBiol2005}.

To verify the hypothesis of decreasing single-track damage probability for protons of high LET, further experimental studies are necessary, as the existing evidence is limited to two survival curves measured under a single experimental setup only. Even though this decrease corresponds to very limited proton ranges in tissues only (approximately 15 $\mu$m or smaller) and will probably not manifest in clinical applications, this finding would be important for understanding the mechanisms underlying the radiobiological effects: For heavier ions, no such complex behaviour has been observed in the studied LET ranges, the damage induction probabilities exhibiting saturation characteristics only (Figure~\ref{fig:ab}). It means that for heavier ions the RBE effects termed often as "overkill" occur as a result of too high energy deposits to individual cells. For protons, on the other hand, this evidence suggests that they are already the effects of single particles on the level of DNA damage formation that get saturated and are responsible for the decrease of biological effectiveness observed experimentally \citep{Belli}.

%%%%%%%%%%%%%%%%%%%%%%%%%%%%%
%%%%%%%%%%%%%%%%%%%%%%%%%%%%%

%%%%%%%%%%%%%%%%%%%%%%%%%%%%%%%%%%%%%%%%%%%%%%%%%%%%%%%%%%%%%%%%%%%%%%%%%%%%%%%%
\subsection{The influence of repair processes}
\label{sec:DiscussionRepair}

\begin{figure}[!htb]
	\centering
		\includegraphics[trim=0cm 0.4cm 0cm 1.2cm, clip, angle=-90, width=0.32\textwidth]{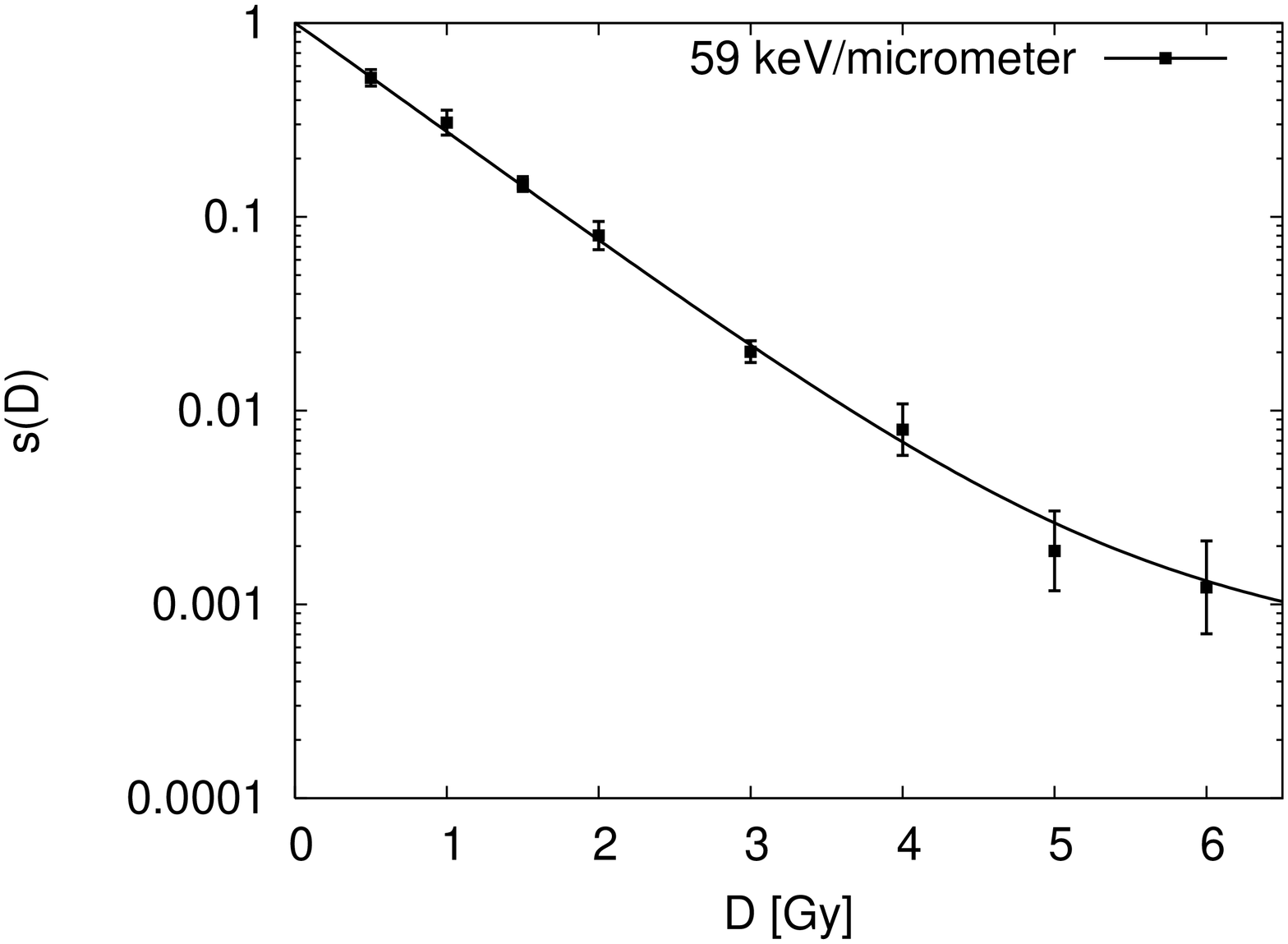}
		\includegraphics[trim=0cm 0.4cm 0cm 1.2cm, clip, angle=-90, width=0.32\textwidth]{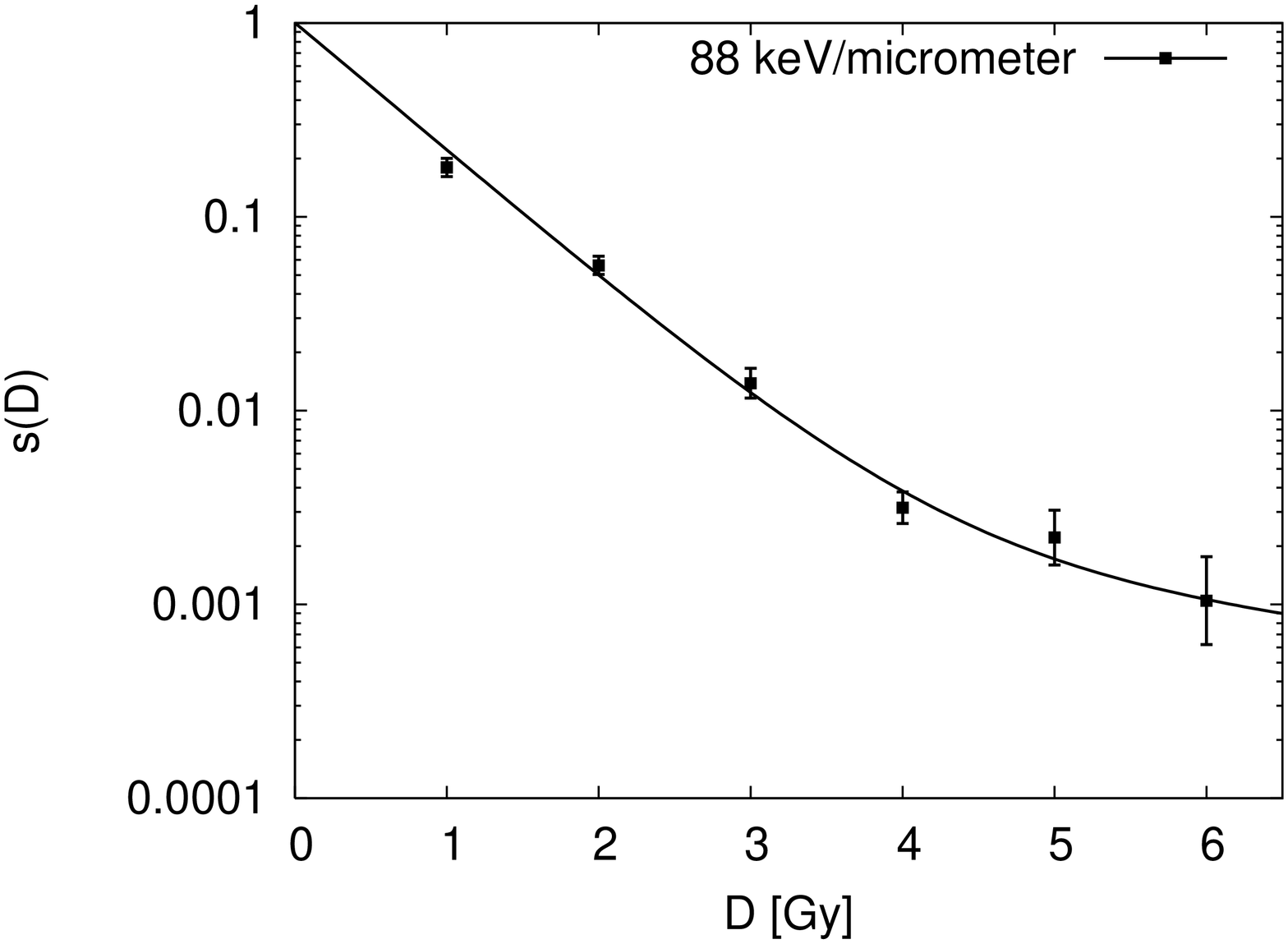}
		\includegraphics[trim=0cm 0.4cm 0cm 1.2cm, clip, angle=-90, width=0.32\textwidth]{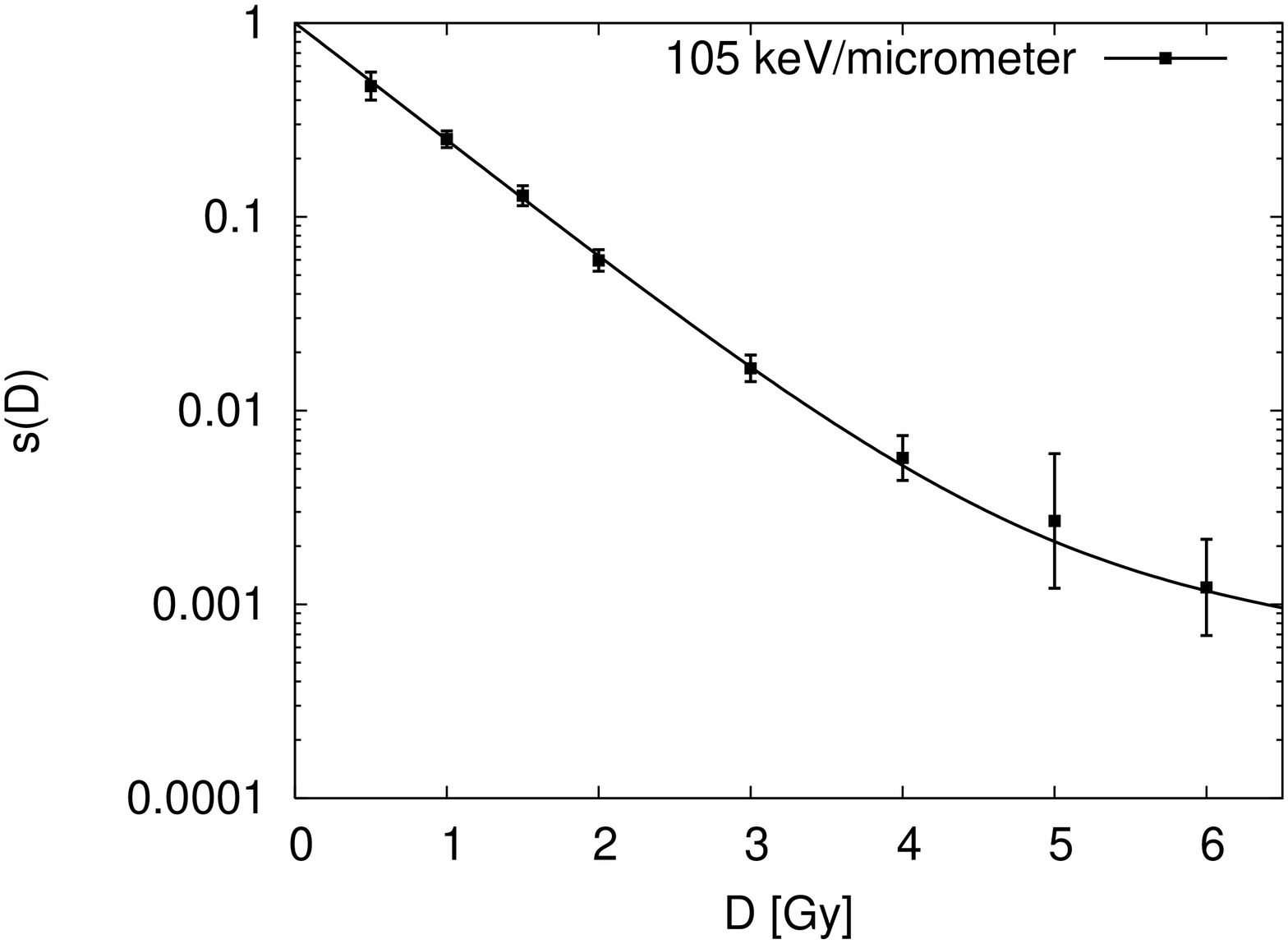}
		\includegraphics[angle=-90, width=0.32\textwidth]{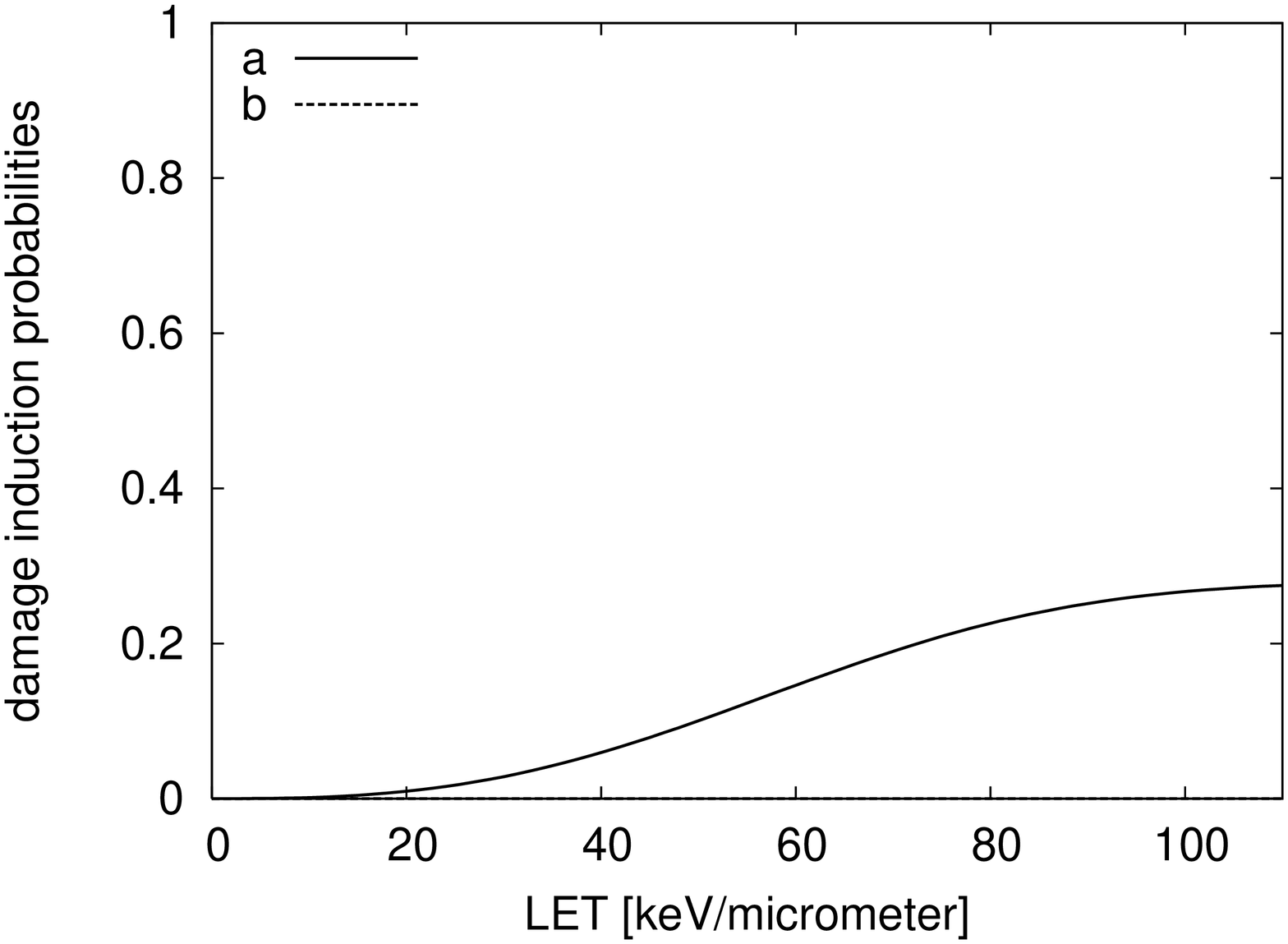}
		\includegraphics[trim=0.7cm 1.2cm 1.3cm 1.6cm, clip, angle=-90, width=0.32\textwidth]{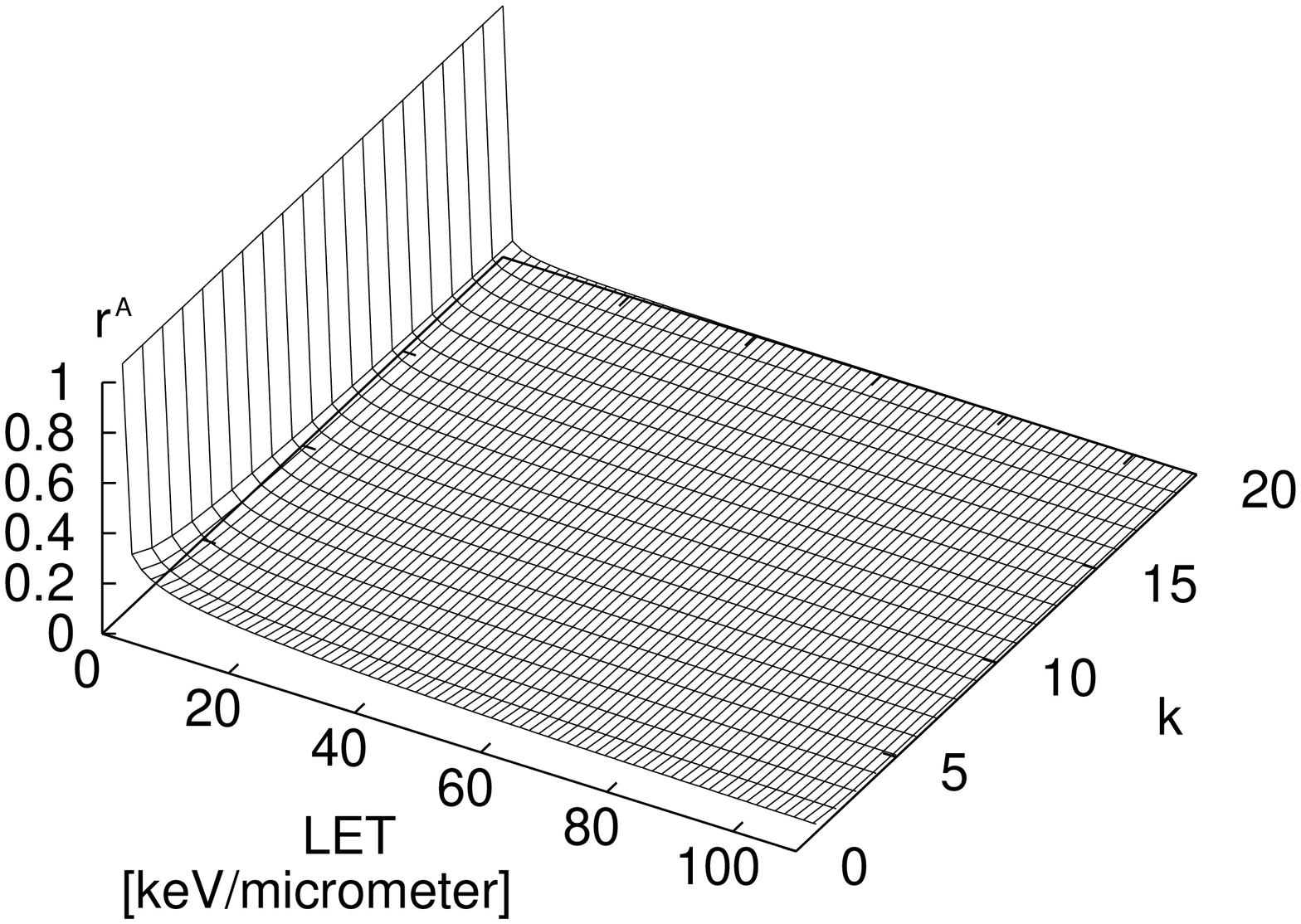}
	\caption{Upper panel: Representation of survival curves for V79 cells irradiated by helium-3 ions (data from \citep{Folkard96}) when repair $r^a$ of single-track damage taken into account. Lower panel: Damage induction and repair probabilities.}
	\label{fig:He-repair}
\end{figure}

As mentioned in the preceding paragraph, the effects of cellular repair processes might be important in representing the outcome of proton irradiation. Also the fits to $^3$He data in the high-dose region might be improved significantly by incorporating the repair probabilities (Figure~\ref{fig:He-repair}). In this region, however, the experimental conditions are very challenging and, consequently, the data in this region have a limited reliability \citep{Folkard96}; therefore, the authors have interpreted the upward-bending of survival curves as artificial effects only. However, the results shown in Figure~\ref{fig:He-repair} demonstrate that similar upward bending might follow as a result of repair processes; compare also the general discussion in \citep{PhysMedBiol2005}. The damage induction probabilities are almost identical with those presented in Figure~\ref{fig:ab} (in both cases, the combined-damage probabilities have been neglected, $b=0$); the survival values at high doses being mainly influenced. Again, additional experimental data would be helpful in solving this issue.

%%%%%%%%%%%%%%%%%%%%%%%%%%%%%
%%%%%%%%%%%%%%%%%%%%%%%%%%%%%

The goodness-of-fits for heavier ions does not increase significantly when the repair processes are taken into account. For the sake of simplicity, only unrepairable damages have, therefore, been considered in the present work (Section~\ref{sec:AnalysisOfExperimentalData}). The induction probabilities for both single-track and combined damages have been estimated on the basis of survival data only. This slightly limits the reliability of estimating the roles of combined-track events, $b$, in the higher-LET regions, whereas the unrepairable single-track damages, $a$, have been assessed reliably in the whole studied LET regions. Somewhat higher values of damage induction probabilities would be obtained if the influence of repair processes were included. A corresponding concept, including a wider class of damages and their repair by the cells, has been used in analyzing radiobiological data for cell lines of different radiosensitivity; the results will be presented and discussed in detail elsewhere \citep{CHO-xrs}.

%%%%%%%%%%%%%%%%%%%%%%%%%%%%%%%%%%%%%%%%%%%%%%%%%%%%%%%%%%%%%%%%%%%%%%%%%%%%%%%%
\subsection{On the interpretation of the damage classification}
\label{sec:DiscussionInterpretation}

Clustered DNA lesions of high complexity (locally multiply damaged sites, complex lesions) are thought to represent significant challenges to cellular repair systems, and to play crucial roles in cell inactivation by ionizing particles \citep{Ward, Goodhead-LMDS, Ottolenghi}. The existing estimates of their yields have been based mostly on Monte Carlo calculations involving track structure simulations and simplified models of DNA and/or chromatin structure. Due to their complexity and computational costs, these approaches have been performed for photons and low-energy particles only. On the other hand, the present approach, based on analyzing survival data, enables to derive the yields of lethal events also for high-$Z$ high-energy ions (Figure~\ref{fig:ab}). By comparing the present results with the mentioned Monte Carlo calculations, mechanistic interpretation of the damage classification used in the present work might be sought, and the fraction of lethal events among different classes of complex lesions can be estimated.

\begin{figure}[!htb]
	\centering
		\includegraphics[trim=0.3cm 0.5cm 0.3cm 1.5cm, clip, angle=0, width=0.80\textwidth]{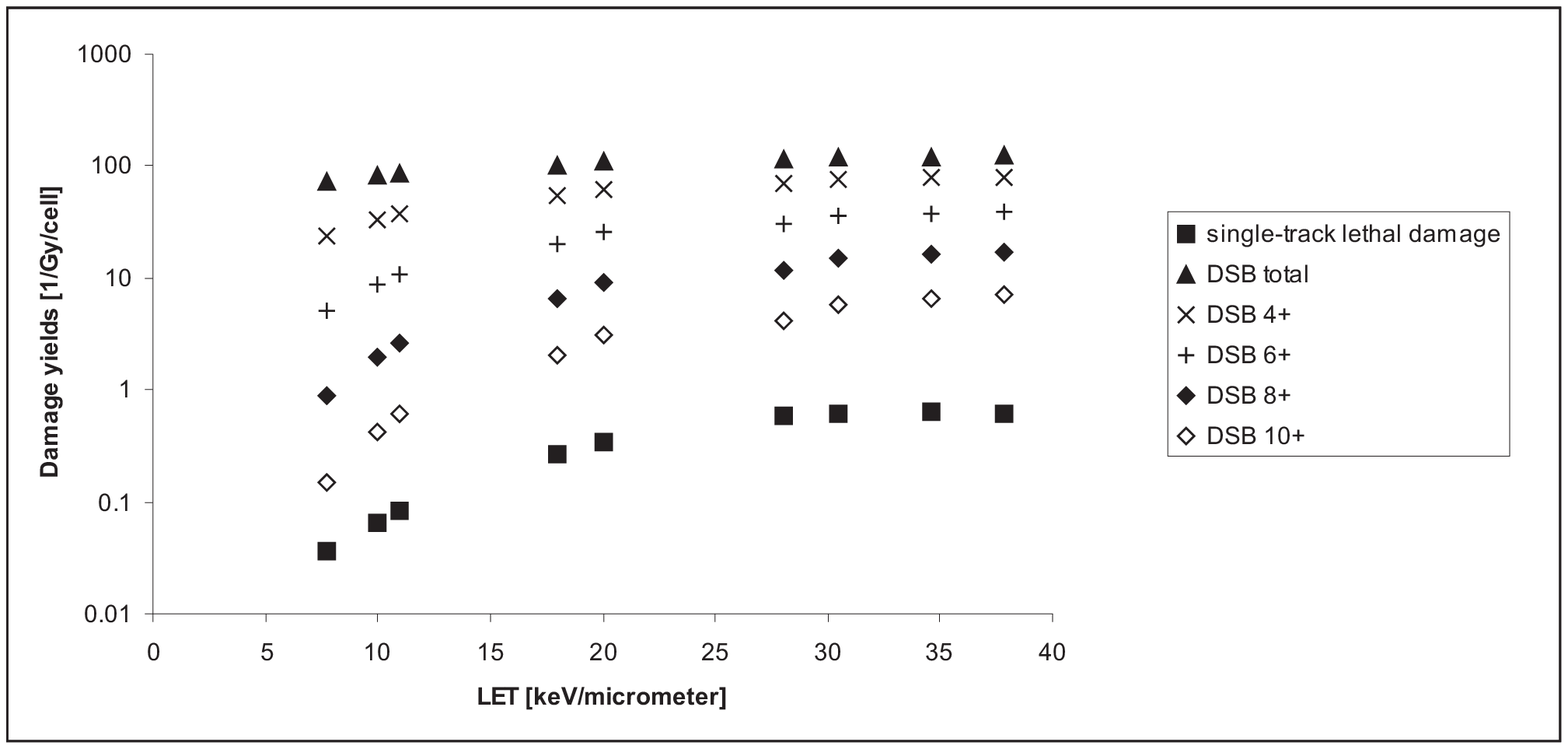}
	\caption{Comparisons of damage yields for protons of different LET values: single-track lethal damage as derived in the present work, and Monte Carlo estimates of total DSB numbers and yields of DSBs of higher complexity, consisting of at least 4, 6, 8, or 10 elementary DNA lesions (DBS 4+, 6+, 8+, 10+, respectively).}
	\label{fig:damageYields}
\end{figure}

\begin{figure}[!htb]
	\centering
		\includegraphics[trim=0.3cm 0.5cm 0.3cm 0.5cm, clip, angle=0, width=0.80\textwidth]{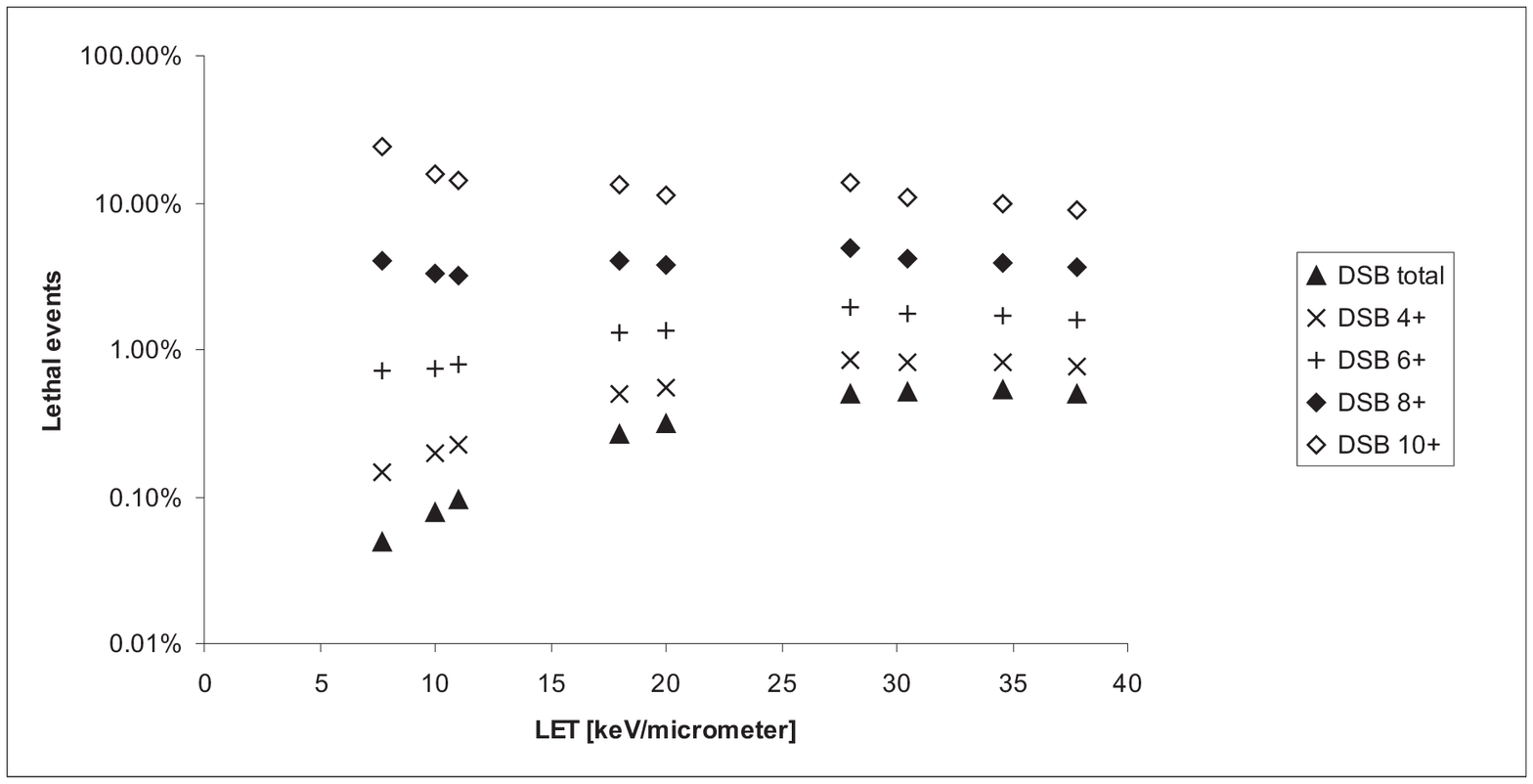}
	\caption{The ratio of lethal events among all DSBs (DSB total) and complex DSBs formed by at least 4, 6, 8, or 10 elementary lesions.}
	\label{fig:lethalRatios}
\end{figure}

As the first step in this direction, the present results for lethal damage induced by protons in V79 cells have been compared to estimates of complex lesions calculated on the basis of a fast Monte Carlo damage simulation model proposed by \citet{Semenenko+Stewart}. In Figure~\ref{fig:damageYields}, the LET-dependent yields of single-track lethal events, $a$, are compared to the total number of double-strand breaks (DSBs) and DSBs composed of at least 4, 6, 8, and 10 elementary lesions (single-strand breaks (SSBs), base damages or abasic sites). This figure indicates that the present results are consistent with the trends predicted by Monte Carlo calculations. In Figure~\ref{fig:lethalRatios}, the ratio between the number of lethal events and DSBs of different complexity is shown, again as a function of LET. Whereas only 0.05~--~0.5\% of all DSBs formed (denoted by DSB total in Figures \ref{fig:damageYields} and~\ref{fig:lethalRatios}) are lethal, this portion increases to approximately 3~--~4\% for DSBs composed of at least 8 elementary lesions (DSB 8+); this number being uniform (LET-independent), i.e.\ being consistent with the hypothesis of a uniform repair probability for these specific damages.

These results indicate that the single-track lethal damages, $a$, might be related to clustered lesions of very high complexity, and correlate best with DSB 8+, containing at least 6 elementary DNA lesions in addition to 2 SSBs forming the DSB. Similarly, the combined damages, $b$, might correspond to pairwise combinations of DSBs formed by different tracks, leading to large-scale chromosomal aberrations; compare e.g.\ \citep{Sachs}. Further damage induction and repair studies are, however, necessary to identify the biophysical nature of lethal events reliably. Furthermore, this comparison has been performed for protons only, given the limitations of existing Monte Carlo approaches. Efforts will therefore be made to extend these calculations to heavier ions and higher energies, using correspondingly simplified Monte Carlo-based models.

%%%%%%%%%%%%%%%%%%%%%%%%%%%%%%%%%%%%%%%%%%%%%%%%%%%%%%%%%%%%%%%%%%%%%%%%%%%%%%%%
\subsection{Applications in hadrontherapy treatment planning}
\label{sec:DiscussionTreatmentPlanning}

The analyzed proton and helium-3 data \citep{Belli, Folkard96, Perris, Goodhead} concern irradiation by low-energy particles only, with ranges in tissue of approx. 10~$\mu$m -- 0.7~mm for protons and 24 -- 74~$\mu$m for $^3$He ions, respectively (Table~\ref{tab:energies}), corresponding to track ends of these particles. On the other hand, data for carbon \citep{Wilma} and oxygen ions \citep{Kiefer} have been gathered over wide energy (and LET) ranges of these particles, corresponding to penetration depths of approx. 45 $\mu$m to 140 mm for carbon ions and 30 $\mu$m to 200 mm for oxygen ions, covering the majority of clinically relevant ranges. Figures~\ref{fig:C} and~\ref{fig:O} illustrate that the simplified probabilistic model enables to represent cell survival over such wide ranges in a systematic manner by using a low number of parameters only (the present analysis involves 6 parameters, $a_0, a_1, a_2$ and $b_0, b_1, b_2$, plus the cross section of cell nuclei, $\sigma$). This is a prerequisite if the model is to be used in treatment planning applications. In the present analysis, cell inactivation effects have been studied for monoenergetic particles as function of their LET value (corresponding to their energy). For treatment planning applications, an adequate physical model, describing the spectra of particle energy and LET values as function of penetration depths and including also energy-loss straggling, lateral scattering and fragmentation processes, has to be used together with the model of biological effects presented here. Such a physical model might be based on detailed Monte Carlo calculations (e.g.\ \citep{SRIM}) or semi-phenomenological parameterization of measured experimental data \citep{TRiP1, Chu}. Calculations of survival as function of penetration depth for several cell lines, using the radiobiological module presented here and a simplified description of physical processes based on SRIM-2003 calculations, have been already performed, showing good agreement with measured data; the results will be presented separately \citep{depth-survival}. To enhance the predictive power of the model, relation between the damage induction probabilities $a, b$ and track structure characteristics of different ions will be sought, with the aim to help in identifying the optimal ion in different clinical situations. Additional systematic experimental data, especially for inactivation induced by light ions (helium to nitrogen), would be helpful in solving this issue, too.

%%%%%%%%%%%%%%%%%%%%%%%%%%%%%%%%%%%%%%%%%%%%%%%%%%%%%%%%%%%%%%%%%%%%%%%%%%%%%%
\section{Conclusion}
\label{sec:Conclusion}
Systematic analysis of published survival data for V79 cells irradiated by light ions has been performed using a simplified scheme of the probabilistic two-stage model. The probabilities of single ions to induce severe damage to DNA have been derived for different ions in dependence on their LET values. In the lower-LET regions, combined damages dominate, while survival at higher LET values is governed by single-particle damages (Figure~\ref{fig:ab}). The present results give quantitative estimates of the increase in damage complexity with increasing atomic number $Z$ and LET value. The analysis of survival data for CHO-K1 cell line shows then the differences in DNA damage induction for different cell lines. 

The derived probabilities of inducing single-track lethal damages have been compared with the Monte Carlo estimates of the yields of complex DNA lesions. Good agreement between these two methodologically different approaches was demonstrated, indicating the biophysical interpretation of lethal events and the importance of damage complexity with respect to its biological consequences.

The results of this work will be used in proposing detailed biology-oriented approaches for hadrontherapy treatment planning. The results may also contribute to understanding the differences between the mechanisms of biological effects of different ions.

%%%%%%%%%%%%%%%%%%%%%%%%%%%%%%%%%%%%%%%%%%%%%%%%%%%%%%%%%%%%%%%%%%%%%%%%

\ack
The author is grateful to Prof R Stewart (Purdue University, USA) for providing him with the results of Monte Carlo simulations of different damage types for low-energy protons and to Prof M Lokaj\'{\i}\v{c}ek for valuable comments on the manuscript. The present work was supported by the grant "Modelling of radiobiological mechanism of protons and light ions in cells and tissues" (Czech Science Foundation, GACR 202/05/2728). Cooperation within the ENLIGHT project ("European Network for LIGht Ion Hadron Therapy", European Commission, QLG1-CT-2002-01574) is acknowledged, too.

%%%%%%%%%%%%%%%%%%%%%%%%%%%%%%%%%%%%%%%%%%%%%%%%%%%%%%%%%%%%%%%%%%%%%%%%

\References

	\bibitem[Belli \etal(1998)]{Belli}
	Belli M, Cera F, Cherubini R, Dalla Vecchia M, Haque AMI, Ianzini F, Moschini G, Sapora O, Simone G, Tabocchini MA, Tiveron P 1998
	RBE-LET relationships for cell inactivation and mutation induced by low energy protons in V79 cells: further results at the LNL facility.
	\emph{Int. J. Radiat. Biol.} \textbf{74} 501--509

	\bibitem[Berger \etal(2000)]{PSTAR}
	Berger MJ, Coursey JS, and Zucker MA 2000
	ESTAR, PSTAR, and ASTAR: Computer Programs for Calculating Stopping-Power and Range Tables for Electrons, Protons, and Helium Ions (version 1.2.2). 
	Available: http://physics.nist.gov/Star. National Institute of Standards and Technology, Gaithersburg, MD.
	Originally published as: Berger MJ, NISTIR 4999, National Institute of Standards and Technology, Gaithersburg, MD (1993).

	\bibitem[Brahme \etal(2001)]{Brahme - Design}
	Brahme A, Lewensohn R, Ringborg U, Amaldi U, Gerardi F, Rossi S 2001
	Design of a centre for biologically optimised light ion therapy in Stockholm.
	\emph{Nucl. Instrum. Meth.} \textbf{B 184} 569-588.

	\bibitem[Cella \etal(2001)]{Lomax2}
	Cella L, Lomax A, Miralbell R 2001 
	Potential role of intensity modulated proton beams in prostate cancer radiotherapy.
	\emph{Int. J. Radiat. Oncol. Biol. Phys.} \textbf{49} 217-223.

	\bibitem[Chu \etal(1993)]{Chu}
	Chu WT, Ludewigt BA, Renner TR 1993
	Instrumentation for treatment of cancer using proton and light-ion beams
	\emph{Rev. Sci. Instrum.} \textbf{64} 2055-2122

	\bibitem[Debus \etal(1998)]{Heidelberg}
	Debus J, Gross KD, Pavlovic M (Eds) 1998
	Proposal for a Dedicated Ion Beam Facility for Cancer Therapy. GSI, Darmstadt.

	\bibitem[Enghardt \etal(2004)]{PET}
	Enghardt W, Parodi K, Crespo P, Fiedler F, Pawelke J, Ponisch F 2004
	Dose quantification from in-beam positron emission tomography 
	\emph{Radiother. Oncol.} \textbf{73} Suppl.\ 2, S96-S98.

	\bibitem[Folkard \etal(1996)]{Folkard96}
	Folkard M, Prise K M, Vojnovic B, Newman H C, Roper M J, Michael B D 1996
	Inactivation of V79 cells by low-energy protons, deuterons and helium-3 ions
	\emph{Int. J. Radiat. Biol.} \textbf{69} 729--738

	\bibitem[Goodhead \etal(1992)]{Goodhead}
	Goodhead DT, Belli M, Mill AJ, Bance DA, Allen LA, Hall SC, Ianzani F, Simone G, Stevens DL, Stretch A, Tabocchini MA, Wilkinson RE	1992
	Direct comparison between protons and alpha-particles of the same LET. I. Irradiation methods and inactivation of asynchronous V79, HeLa and C3H 10T1/2 cells
	\emph{Int. J. Radiat. Biol.} \textbf{61} 611--624

	\bibitem[Goodhead(1994)]{Goodhead-LMDS}
	Goodhead DT 1994 
	Initial events in the cellular effects of ionizing-radiations - clustered damage in DNA
	\emph{Int. J. Radiat. Biol.} \textbf{65} 7-17

	\bibitem[Hrom\v{c}\'{\i}kov\'{a} \etal(2005)]{CHO-xrs}
	Hrom\v{c}\'{\i}kov\'{a} H, Kundr\'{a}t P, Lokaj\'{\i}\v{c}ek M 2005
	Analysis of damage induction and repair processes after carbon irradiation in different cell lines
	(in preparation)

	\bibitem[James(1994)]{MINUIT}
	James F 1994
	MINUIT Minimization package - Reference Manual
	\emph{CERN Program Library Long Writeup} D506, CERN Geneva

	\bibitem[Kagawa \etal (2002)]{Kagawa 2002}
	Kagawa K, Murakami M, Hishikawa Y, Abe M, Akagi T, Yanou T, Kagiya G, Furusawa Y, Ando K, Nojima K, Aoki M, Kanai T 2002
	Preclinical biological assessment of proton and carbon ion beams at Hyogo Ion Beam Medical Center. 
	\emph{Int. J. Radiat. Oncol. Biol. Phys.} \textbf{54} 928--938

	\bibitem[Kanai \etal (1997)]{Kanai 1997}
	Kanai T, Furusawa Y, Fukutsu K, Itsukaichi H, Eguchi-Kasai K, Ohara H 1997
	Irradiation of Mixed Beam and Design of Spread-out Bragg Peak for Heavy-Ion Radiotherapy.
	\emph{Radiat. Res.} \textbf{147} 78--85

	\bibitem[Kanai \etal (1999)]{Kanai 1999}
	Kanai T, Endo M, Minohara S, Miyahara N, Koyama-Ito H, Tomura H, Matsufuji N, Futami Y, Fukumura A, Hiraoka T, Furusawa Y, Ando K, Suzuki M, Soga F, Kawachi K 1999
	Biophysical characteristics of HIMAC clinical irradiation system for heavy-ion radiation therapy.
	\emph{Int. J. Radiat. Oncol. Biol. Phys.} \textbf{44} 201--210

	\bibitem[Kr\"{a}mer \etal(2000)]{TRiP1}
	Kr\"{a}mer M, J\"{a}kel O, Haberer T, Kraft G, Schardt D, Weber U 2000
	Treatment planning for heavy-ion radiotherapy: physical beam model and dose optimization.
	\emph{Phys. Med. Biol.} \textbf{45} 3299-3317.

	\bibitem[Kr\"{a}mer and Scholz(2000)]{TRiP2}
	Kr\"{a}mer M, Scholz M 2000
	Treatment planning for heavy-ion radiotherapy: calculation and optimization of biologically effective dose
	\emph{Phys. Med. Biol.} \textbf{45} 3319-3330

	\bibitem[Krengli and Orecchia(2004)]{CNAO}
	Krengli M, Orecchia R 2004
	Medical aspects of the National Centre for Oncological Hadrontherapy (CNAO - Centro Nazionale Adroterapia Oncologica) in Italy.
	\emph{Radiother. Oncol.} \textbf{73} Suppl.\ 2, S21-S23.

	\bibitem[Kundr\'{a}t \etal(2005)]{PhysMedBiol2005}
	Kundr\'{a}t P, Lokaj\'{\i}\v{c}ek M, Hrom\v{c}\'{\i}kov\'{a} H 2005
	Probabilistic two-stage model of cell inactivation by ionizing particles 
	\emph{Phys. Med. Biol.} \textbf{50} 1433-1447

	\bibitem[Kundr\'{a}t(2005)]{depth-survival}
	Kundr\'{a}t P 2005
	Towards biology-oriented hadrontherapy treatment planning 
	(in preparation)

	\bibitem[Mayer \etal(2004)]{MedAustron}
	Mayer R, Mock U, Jager R, Potter R, Vutuc C, Eiter H, Krugmann K, Hammer J, Hirn B, Hawliczek R, Knocke-Abulesz TH, Lukas P, Nechville E, Pakisch B, Papauschek M, Wolfgang R, Rhomberg W, Sabitzer H, Schratter-Sehn A, Felix S, Wedrich I, Auberger T 2004
	Epidemiological aspects of hadron therapy: A prospective nationwide study of the Austrian project MedAustron and the Austrian Society of Radiooncology (OEGRO).
	\emph{Radiother. Oncol.} \textbf{73} Suppl.\ 2, S24-S28.

	\bibitem[Ottolenghi \etal(1997)]{Ottolenghi}
	Ottolenghi A, Merzagora M, Paretzke HG 1997
	DNA complex lesions induced by protons and alpha-particles: Track structure characteristics determining linear energy transfer and particle type dependence 
	\emph{Radiat. Environ. Biophys.} \textbf{36} 97-103

	\bibitem[Perris \etal(1986)]{Perris}
	Perris A, Pialoglou P, Katsanos AA, Sideris EG 1986
	Biological effectiveness of low-energy protons. 1. Survival of Chinese-hamster cells
	\emph{Int. J. Radiat. Biol.} \textbf{50} 1093--1101

	\bibitem[Sachs and Brenner(1993)]{Sachs}
	Sachs RK, Brenner DJ 1993
	Effect of LET on chromosomal aberration yields. 1. Do long-lived, exchange-prone double-strand breaks play a role 
	\emph{Int. J. Radiat. Biol.} \textbf{64} 677-688

	\bibitem[Schall \etal(1996)]{fragmentation}
	Schall I, Schardt D, Geissel H, Irnich H, Kankeleit E, Kraft G, Magel A, Mohar MF, Munzenberg G, Nickel F, Scheidenberger C, Schwab W 1996
	Charge-changing nuclear reactions of relativistic light-ion beams ($5 \leq Z \leq 10$) passing through thick absorbers 
	\emph{Nucl. Instrum. Meth.} \textbf{B 117} 221-234 

	\bibitem[Scholz and Kraft(1995)]{Kraft-Scholz-50}
	Scholz M, Kraft G 1995
	Track structure and the calculation of biological effects of heavy charged-particles
	\emph{Adv. Space Res.} \textbf{18} 5-14

	\bibitem[Schulz-Ertner \etal(2002)]{chordomas}
	Schulz-Ertner D, Haberer T, Jakel O, Thilmann C, Kramer M, Enghardt W, Kraft G, Wannenmacher M, Debus JR 2002
	Radiotherapy for chordomas and low-grade chondrosarcomas of the skull base with carbon ions 
	\emph{Int. J. Radiat. Oncol. Biol. Phys.} \textbf{53} 36-42.

	\bibitem[Semenenko and Stewart (2004)]{Semenenko+Stewart}
	Semenenko VA, Stewart RD 2004
	A fast Monte Carlo algorithm to simulate the spectrum of DNA damages formed by ionizing radiation 
	\emph{Radiat. Res.} \textbf{161} 451-457

	\bibitem[Sisterson(2005)]{Particles}
	Sisterson J (Ed) (2005)
	\emph{Particles Newsletter} 36.
	http://ptcog.web.psi.ch/archive.html

	\bibitem[Stoll \etal(1995)]{Kiefer}
	Stoll U, Schmidt A, Schneider E, Kiefer J 1995
	Killing and mutation of Chinese-hamster V79 cells exposed to accelerated oxygen and neon ions.
	\emph{Radiat. Res.} \textbf{142} 288-294

	\bibitem[Tsujii \etal(2004)]{Tsujii - PTCOG Paris 2004}
	Tsujii H, Mizoe J, Kamada T, Baba M, Kato S, Kato H, Tsuji H, Yamada S, Yasuda S, Ohno T, Yanagi T, Hasegawa A, Sugawara T, Ezawa H, Kandatsu S, Yoshikawa K, Kishimoto R, Miyamoto T 2004
	Overview of clinical experiences on carbon ion radiotherapy at NIRS.
	\emph{Radiother. Oncol.} \textbf{73} Suppl.\ 2, S41-S49

	\bibitem[Wambersie \etal(2004)]{ion-rationale}
	Wambersie A, Hendry J, Gueulette J, Gahbauer R, Potter D, Gregoire V 2004
	Radiobiological rationale and patient selection for high-LET radiation in cancer therapy 
	\emph{Radiother. Oncol.} \textbf{73} Suppl.\ 2, S1-S14.

	\bibitem[Ward (1985)]{Ward}
	Ward JF 1985
	Biochemistry of DNA lesions
	\emph{Radiat. Res.} \textbf{104} S103-S111

	\bibitem[Weyrather \etal(1999)]{Wilma}
	Weyrather WK, Ritter S, Scholz M, Kraft G 1999
	RBE for carbon track-segment irradiation in cell lines of differing repair capacity
	\emph{Int. J. Radiat. Biol.} \textbf{75} 1357--1364

	\bibitem[Wilson (1946)]{Wilson}
	Wilson RR 1946 
	Radiological use of fast protons. 
	\emph{Radiology} \textbf{47} 487-491

	\bibitem[Ziegler(2004)]{SRIM}
	Ziegler JF 2004
	SRIM-2003
	\emph{Nucl. Instrum. Meth.} \textbf{B 219} 1027-1036 

	\bibitem[Zurlo \etal(2000)]{Lomax1}
	Zurlo A, Lomax A, Hoess A, Bortfeld T, Russo M, Goitein G, Valentini V, Marucci L, Capparella R, Loasses A 2000
	The role of proton therapy in the treatment of large irradiation volumes: A comparative planning study of pancreatic and biliary tumors. 
	\emph{Int. J. Radiat. Oncol. Biol. Phys.} \textbf{48} 277-288.

\endrefs

\end{document}